\documentclass[journal=jctcce,manuscript=article,hyphens,maxauthors=60]{achemso}
\usepackage[T1]{fontenc}

\usepackage[version=3]{mhchem} 
\usepackage{multirow}
\usepackage[table,xcdraw]{xcolor}
\usepackage{booktabs}  
\usepackage{array}      
\usepackage{amsmath}    
\usepackage{float} 
\usepackage{colortbl}
\usepackage{braket}
\usepackage{longtable}
\usepackage{rotating}
\usepackage{graphicx}
\usepackage{threeparttable}
\usepackage{subcaption}
\usepackage{bm}
\usepackage{nicefrac}

\newcommand{\Lower}[1]{\smash{\lower 1.5ex \hbox{#1}}}
\newcommand{\Tp}{{{T}_2^{\rm{pCCD}}}}
\usepackage{soul}  
\usepackage{xcolor}  
\sethlcolor{yellow}  
\soulregister\cite7

\usepackage{xcolor}

\author{Seyedehdelaram Jahani}
\affiliation{Institute of Physics, Faculty of Physics, Astronomy, and Informatics, Nicolaus Copernicus University in Toruń, Grudziądzka 5, 87-100 Toru\'{n}, Poland.}
\author{Katharina Boguslawski}
\email{k.boguslawski@fizyka.umk.pl}
\affiliation{Institute of Physics, Faculty of Physics, Astronomy, and Informatics, Nicolaus Copernicus University in Toruń, Grudziądzka 5, 87-100 Toru\'{n}, Poland.}
\author{Pawe\l{} Tecmer}
\affiliation{Institute of Physics, Faculty of Physics, Astronomy, and Informatics, Nicolaus Copernicus University in Toruń, Grudziądzka 5, 87-100 Toru\'{n}, Poland.}
 \email{ptecmer@fizyka.umk.pl}

\title[Ionization Potentials at Mean-Field Computational Cost]
{Ionization Potentials at Mean-Field Computational Cost: The Extended Koopmans' Framework for pCCD}

\abbreviations{CC, EOM-CC, HF, oo-pCCD, KT, EKT, IP}

\begin{document}

\begin{tocentry}


\includegraphics[width=8.25cm,height=4.45cm]{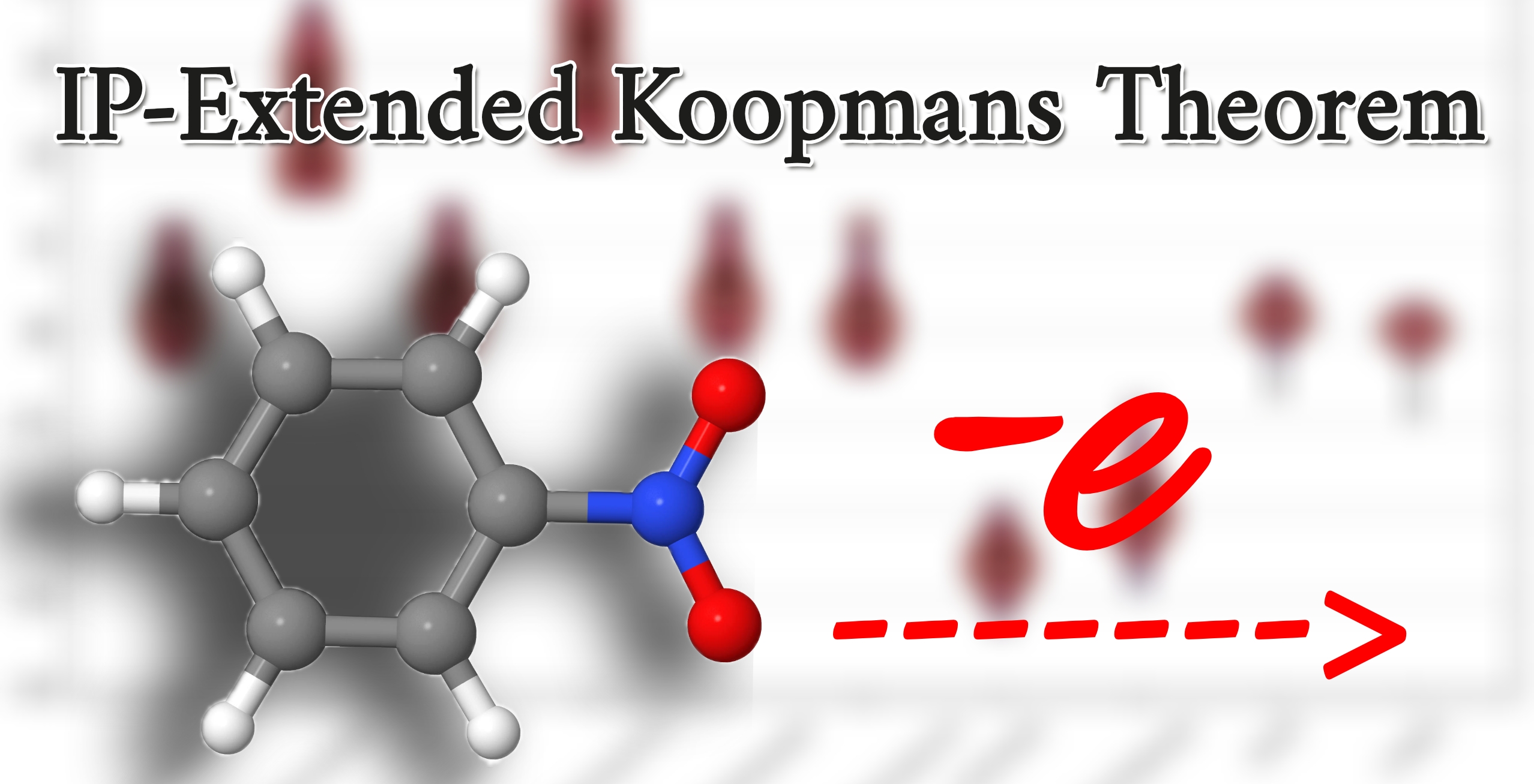}

\end{tocentry}

\begin{abstract}
We introduce a mean-field-like computational model for calculating ionization potentials (IPs) based on the pair Coupled Cluster Doubles (pCCD) wavefunction. 
Specifically, our model combines the extended Koopmans' theorem (EKT) with the advantages of a variationally orbital-optimized (oo)-pCCD ansatz. 
The computational cost of the EKT(pCCD) method is negligible ($\mathcal{O}(N^3)$) as the response 1- and 2-particle reduced density matrices used to construct the generalized Fock matrix are readily available after an oo-pCCD calculation.
We benchmarked our new computational model for IPs of atoms, small molecules, and a set of organic acceptor molecules against experimental and theoretical reference data. 
The EKT(pCCD) model significantly improves upon the modified Koopmans' approach [J. Chem. Phys. 162, 184110 (2025)], and the obtained IPs are comparable to those of computationally more expensive IP-EOM-pCCD-based models, approaching CCSD(T) reference values (with a mean error of 0.05 eV).
Most importantly, the EKT(pCCD) approach is almost independent of the basis set size, and reliable IPs are already obtained with small basis sets. 
\end{abstract}


\section{Introduction} \label{sec:Introduction}
Recent decades have seen intensified efforts to develop efficient and reliable organic solar cells as flexible, low-cost alternatives to silicon-based photovoltaics.~\cite{opv-mat-today-2012, pi-organic-electronics-chem-mater-2014, non-fullerene-opv-nat-commun-2021}
Organic photovoltaic (OPV) devices can now exceed 19$\%$ efficiency through optimized materials design and device engineering in laboratory conditions.~\cite{19-efficiency-single-junction-osc-nat-mat-2022}
Key to this progress is accurate control of electronic properties, particularly the energy offset between a donor’s highest occupied molecular orbital (HOMO) and an acceptor’s lowest unoccupied molecular orbital (LUMO) and the first optically active excited state.~\cite{bhj-performance-chem-soc-rev-2011,polymer-solar-cells-nat-photonics-2012}
Accurate prediction of HOMO-LUMO gaps (also known as charge gaps) and optical gaps remains critical for optimizing charge separation, stability, and device performance.~\cite{molecular-understanding-osc-acc-chem-res-2009}
Quantum chemistry methods, particularly density functional theory (DFT), form the backbone of contemporary research on organic solar cells\mbox{~\cite{risko-qc-perspective-chem-sci-2011}}.
Although DFT is widely used for electronic structure calculations in organic photovoltaics (OPV), direct predictions of ionization potentials (IPs) and electron affinities (EAs) from Kohn--Sham orbital energies are problematic with approximate functionals: IPs are typically inaccurate (except for the exact functional, 
where $-\epsilon_{\rm HOMO}$ equals the IP, albeit subject to caveats regarding the asymptotic decay of the exchange--correlation potential\mbox{~\cite{fscc-spin-free-triatomics-pccp-2011, pccd-delaram-rsc-adv-2023}}), while EAs derived from orbital energies generally correspond to excitations rather than true affinities.\mbox{~\cite{filippi-dft-jcp-1997,savin-Kohn-Sham-cpl-1998,tozer-Kohn-Sham-jcp-1998,tozer-dft-jpca-2005,dft-anions-frank-jensen-jctc-2010,dft-advances-in-organic-electronics-acr-2014,19-efficiency-single-junction-osc-nat-mat-2022,pccd-delaram-rsc-adv-2023}}

From a theoretical standpoint, accurately determining IPs and EAs has long been a significant challenge in quantum chemistry.
The IP offers insight into a system's reactivity by indicating how easily an electron can be removed from a molecule, thereby quantifying the molecule's tendency to form a positively charged ion or donate an electron.
On the contrary, EA describes the tendency to form an anion and accept an electron.
While well-established computational techniques such as electron propagator theory (EPT)~\cite{propagator-theory-book-aqc-1981, propagators-in-quantum-chemistry-book-2004, electron-propagator-wires-2012}, ionization potential equation-of-motion coupled cluster (IP-EOM-CC) methods~\cite{ip-eom-cc-ijqc-1992, ip-eom-cc-stanton-jcp-1994, ip-eom-cc-stanton-jcp-1999, ip-eom-pccd-chem-comm-2021, tailored-ip-eom-pccd-jctc-2024}, and electron affinity equation-of-motion coupled cluster (EA-EOM-CC) methods~\cite{sacci-naktsuji-cpl-1991, ea-eom-cc-marcel-jcp-1995, ea-eom-ccsdt-jcp-2003} deliver reliable results, they frequently incur substantial computational expenses.
Related examples include low-scaling methods based on second-order many-body perturbation theory approximations~\cite{scs-delta-mp2-jctc-2018}, Green's wavefunction (GW) algorithms,~\cite{ip-ea-gw-benchmark-jctc-2021} the second-order algebraic diagrammatic construction (ADC(2)).~\cite{ip-cc2-adc2-jctc-2022}
An alternative to compute IPs (and to some extent EAs) is to use the extended Koopmans' theorem (EKT)~\cite{ekt-from-2-rdm-ijqc-1974, ekt-ip-jcp-1975, ekt-jcp-1977, ekt-morrison-jcp-1992, ekt-tca-1994, ekt-cioslowski-jcp-1997, ekt-validity-pernal-jcp-2001, ekt-exact-ip-jcp-2009, ekt-validity-ernzerhof-jctc-2009, ekt-review-advances-in-quantum-chemistry-2022,ekt-ip-dmft-cpl-2005, ekt-pnos-jcp-2012, ekt-ip-bozkaya-jcp-2013, ekt-ea-bozkaya-jctc-2014, ekt-mp2-jcc-2020, ekt-casscf-jcp-2021, ekt-casscf-jcp-2024} on top a correlated wavefunction. 
The original formulation of EKT that relies on variationally determined density matrices~\cite{ekt-from-2-rdm-ijqc-1974} has recently been extended to orbital-optimized coupled cluster and perturbation theory methods by Bozkaya.~\cite{ekt-ip-bozkaya-jcp-2013, ekt-ea-bozkaya-jctc-2014} 
The EKT formulation introduced by Cioslowski \textit{et al.}~\cite{ekt-cioslowski-jcp-1997} enables the integration of EKT with any wavefunction-based approach that supports analytic gradient computations. 

Reliable and efficient computational methods are essential for modeling the electronic structures and properties of organic electronic material building blocks.
Among these, pair Coupled Cluster Doubles (pCCD)~\cite{ap1rog-piotrus-jctc-2013, oo-ap1rog-prb-2014, ap1rog-non-variational-orbital-optimization-jctc-2014, tamar-pccd-jcp-2014, ap1rog-orbital-optimization-piotrus-mol-phys-2014, ap1rog-orbital-energies-jctc-2015} and related geminal-based approaches~\cite{ccvb-pccp-2011,ccvb-jctc-2017,lr-gvb-dalton-hapka-jcp-2022, pj-seniority-gs-es-rg-jcp-2022, pj-density-matrices-jcp-2022, pj-rmds-rg-jcp-2022, ramon-cc-inspired-geminals-jcp-2024, block-correlated-gvb-eom-jcpl-2025, seniority-zero-wavefunction-ayers-tca-2025} have demonstrated strong potential to overcome limitations of traditional methods.~\cite{pccd-delaram-rsc-adv-2023, pccd-perspective-jpcl-2023, pccd-dipole-moments-jctc-2024, pccd-expectation-value-1dm-jpca-2025,ea-eom-fpccd-jctc-2025,lena-ct-jctc-2025,dip-fpccsd-jctc-2025,ea-eom-pccd-jpca-2024,pccd-ip-ea-mod-koopmans-jcp-2025}
Simplified coupled-cluster models, such as the pCCD method, offer a cost-effective and reliable framework for describing quasi-degenerate and strongly correlated systems~\cite{ap1rog-piotrus-jctc-2013,diatomics-oo-ap1rog-jpca-2014, ap1rog-orbital-optimization-piotrus-mol-phys-2014, oo-ap1rog-prb-2014, ap1rog-singlet-gs-actinides-pccp-2015, ap1rog-orbital-energies-jctc-2015, pccd-correlation-analysis-prb-2016, pccd-yb2-ijqc-2019}.
When combined with an efficient orbital optimization (oo) protocol~\cite{oo-ap1rog-prb-2014,ap1rog-non-variational-orbital-optimization-jctc-2014}, the oo-pCCD method achieves size consistency. 
In this work, we implement the EKT approach within the oo-pCCD framework~\cite{oo-ap1rog-prb-2014, ap1rog-non-variational-orbital-optimization-jctc-2014} to compute reliable IPs at low computational cost (mean-field-like). 
As any other orbital-optimized CC methods~\cite{bozkaya-oo-ccd-jcp-2011,bozkaya-oo-coupled-electron-jcp-2013,bozkaya-oo-lccd-lambda-equations-pccp-2016,bozkaya-symmetric-oo-ccd-jcp-2012}, oo-pCCD is ideally suited for EKT calculations as it offers direct access to reduced density matrices (RDMs) and the generalized Fock matrix (GFM).
Moreover, the N-representability condition for the one-particle reduced density matrix (1-RDM) from oo-pCCD is satisfied, and orbital relaxation effects are accounted for.~\cite{pccd-dipole-moments-jctc-2024} 

Most importantly, oo-pCCD\cite{tamar-pccd-jcp-2014, oo-ap1rog-prb-2014} has a more favorable computational scaling than traditional coupled-cluster methods (CC), such as CCD and CCSD.\mbox{\cite{bozkaya-oo-ccd-jcp-2011,bozkaya-symmetric-oo-ccd-jcp-2012,bozkaya-oo-lccd-lambda-equations-pccp-2016}}
A single pCCD evaluation formally scales as $\mathcal{O}(o^2 v^2)$ (where $o$ and $v$ denote the number of occupied and virtual orbitals, respectively), which can be reduced to $\mathcal{O}(o v^2)$ scaling of the effective particle-particle ladder term.
Nonetheless, the four-index transformation represents the bottleneck operation, deteriorating the scaling to $\mathcal{O}(N^5)$ (where $N$ is the number of basis functions) or $\mathcal{O}(N^4)$ if Cholesky-decomposed electron repulsion integrals\mbox{\cite{pedersen-cholesky-cc-jcp-2007,cholesky-review-2011}} are used.\mbox{~\cite{oo-ap1rog-prb-2014, ap1rog-non-variational-orbital-optimization-jctc-2014}}
The orbital optimization protocol introduces additional cost due to the construction of the orbital gradient and its intermediates, leading to approximately $\mathcal{O}(N^3)$ or $\mathcal{O}(N^4)$ (in the case of Cholesky-decomposed electron repulsion integrals) cost.
Thus, the orbital-optimized variant of pCCD is dominated by $\mathcal{O}(N^4)$ scaling at best.

This work is structured as follows: Section~\ref{sec:Theory} offers a concise overview of the theoretical models under investigation, followed by a detailed description of the computational methods in Section~\ref{sec:comput-det}. Section~\ref{sec:results} summarizes the numerical results along with a statistical analysis.
Concluding remarks are presented in Section~\ref{sec:conclusions}.
\section{Theory}\label{sec:Theory}
\subsection{The pCCD ansatz}
An alternative conceptual approach to capturing strong electron correlation is to embed electron correlation effects directly within the electronic wavefunction through the use of two-electron functions, commonly referred to as geminals. When these geminals are constrained to singlet (two-electron) functions, the pair-excitation function, in its natural form~\cite{surjan-geminals-springer-1999,surjan-geminals-jmathchem-2012}, can be written as
\begin{equation}
\Psi_i^{\dagger} = \sum_{p=1}^{M_i} c_p^i\, a_p^{\dagger} a_{\bar{p}}^{\dagger}
\label{eq:geminal-creation-operator}
\end{equation}
Here, $a_{p}^\dagger$ and $a_{\bar{p}}^\dagger$ denote the electron creation operators for spin-up ($p$) and spin-down ($\bar{p}$) electrons in orbital $p$, respectively. The matrix $(c_{p}^{i})$, of dimension $P \times M$, serves as the geminal coefficient matrix relating the underlying one-particle orbitals to the two-particle geminal creation operators $\Psi_i^\dagger$. According to Eq.~\eqref{eq:geminal-creation-operator}, each geminal $\Psi_i^\dagger$ is formed from orbitals restricted to (possibly disjoint) subspaces $M_i$. The resulting geminal-based wavefunction can therefore be written as

\begin{equation}
\lvert {\rm Geminal} \rangle = \prod_{i}^{P} \Psi_i^{\dagger} \lvert 0 \rangle
\label{eq:geminal-wavefunction}
\end{equation}
Here, $P$ denotes the number of electron pairs, and $|0\rangle$ represents the vacuum state with respect to the geminal creation operators.

Among practical geminal methods, the antisymmetric product of 1-reference-orbital geminals (AP1roG)~\cite{ap1rog-piotrus-jctc-2013, oo-ap1rog-prb-2014} stands out as a promising alternative in large-scale modeling.~\cite{pccd-delaram-rsc-adv-2023,pccd-perspective-jpcl-2023, ea-eom-pccd-jpca-2024, lena-ct-jctc-2025, dip-fpccsd-jctc-2025}
AP1roG can be reformulated exactly as a fully general pair-Coupled-Cluster Doubles (pCCD) wavefunction~\cite{ap1rog-piotrus-jctc-2013, oo-ap1rog-prb-2014, ps2-ap1rog-jcp-2014, ap1rog-non-variational-orbital-optimization-jctc-2014,tamar-pccd-jcp-2014, pccp-geminal-review-pccp-2022}, expressed as
\begin{equation}
\label{eq:ap1rog}
\left|\Psi_{\rm{pCCD}}\right\rangle = 
\exp \left( \sum_{i=1}^{P} \sum_{a=P+1}^{K} c_{i}^{a} \, a^\dagger_{a} a^\dagger_{\bar a} a_{\bar i} a_{i} \right) \left|\Phi_0\right\rangle =e^{\Tp} 
| \Phi_0 \rangle,
\end{equation}
where $a^\dagger_{p}$ and $a_{p}$ are the fermionic creation and annihilation operators, respectively, for $\alpha$ ($p$) and $\beta$ ($\bar p$) electrons.
The state $\left|\Phi_0\right\rangle$ represents an independent-particle wavefunction, often chosen as the Hartree--Fock (HF) determinant, and $\Tp$ is the cluster operator containing electron pair excitations.
The indices $i$ and $a$ label occupied and virtual (spatial) orbitals relative to $\left|\Phi_0\right\rangle$ (in the following, we work with restricted orbitals only).
Here, $P$ and $K$ denote the number of electron pairs ($P = N/2$, where $N$ is the total electron count) and the total number of (spatial) orbitals.
The coefficients $\{ c_{i}^{a} \}$ correspond to the geminal amplitudes.
This ansatz for the wavefunction is size-extensive.~\cite{oo-ap1rog-prb-2014, ap1rog-orbital-optimization-piotrus-mol-phys-2014, tamar-pccd-jcp-2014, ps2-ap1rog-jcp-2014, kasia-ap1rog-jctc-2014, state-specific-oopccd-jctc-2021,ap1rog-piotrus-jctc-2013}. Moreover, the pCCD molecular orbitals, which serve to define the reference determinant in Eq.~\eqref{eq:ap1rog}, are generally optimized through a variational orbital optimization procedure~\cite{oo-ap1rog-prb-2014, ap1rog-orbital-optimization-piotrus-mol-phys-2014} or alternative procedures.~\cite{ps2-ap1rog-jcp-2014, ap1rog-non-variational-orbital-optimization-jctc-2014}

In oo-pCCD, the orbitals are selected to minimize the pCCD energy functional subject to the constraint that the pCCD coefficient equations are satisfied~\cite{ap1rog-piotrus-jctc-2013,kasia-ap1rog-jctc-2014, ap1rog-orbital-energies-jctc-2015}. Under intermediate normalization, the energy Lagrangian takes the following form
\begin{equation}
\label{eq:energy_lagrangian_AP1roG}
\mathcal{L} = \langle \Phi_0 | e^{-\Tp} e^{\kappa} H e^{-\kappa} e^{\Tp} | \Phi_0 \rangle 
            + \sum_{i,a} \lambda_i^a \left( \langle \Phi_{i{\bar{i}}}^{a{\bar{a}}} | e^{-\Tp} e^{\kappa} H e^{-\kappa} e^{\Tp} | \Phi_0 \rangle \right),
\end{equation}
where $\{\lambda_i^a\}$ denote the Lagrange multipliers and the Hamiltonian is explicitly expressed in the rotated orbital basis, with $\kappa$ denoting the generator of the orbital rotations,
\begin{equation}
\kappa = \sum_{p > q} \kappa_{pq} \big( a_p^\dagger a_q - a_q^\dagger a_p \big)
\label{eq:orbital-rotation-operator-1}
\end{equation}
and includes all nonredundant orbital rotations within the occupied--occupied, occupied--virtual, and virtual--virtual orbital subspaces.~\cite{oo-ap1rog-prb-2014}
For spatial orbitals, we can rewrite $\kappa$ as
\begin{equation}
\kappa = \sum_{p > q} \kappa_{pq} \big( E_{pq} - E_{qp} \big),
\label{eq:orbital-rotation-operator}
\end{equation}
where $E_{pq} = a_p^\dagger a_q + {a}_{\bar p}^\dagger {a}_{\bar q}$ represents the singlet excitation operator, with $p$ and $q$ running over all active orbitals, that is occupied and virtual ones.
The determinant $|\Phi^{\bar{a}\bar{a}}_{\bar{i}\bar{i}}\rangle = a^\dagger_a a^\dagger_{\bar{a}} a_{\bar{i}} a_i |\Phi_0\rangle$ corresponds to a pair-excited Slater determinant.

Imposing the partial derivative of $\mathcal{L}$ with respect to the Lagrange multipliers $\{\lambda_i^a\}$ to be stationary leads to the conventional equations for the geminal coefficients $\{c_i^a\}$ (evaluated for the current set of orbitals $\kappa=0$)
\begin{equation}
\label{eq:stationary_condition_geminal_coeff}
\frac{\partial \mathcal{L}}{\partial \lambda_i^a} \Big\vert_{\kappa=0}
= \langle \Phi_{i{\bar{i}}}^{a{\bar{a}}} | e^{-\Tp} H e^{\Tp} | \Phi_0 \rangle = 0
\end{equation}
The stationary condition of $\mathcal{L}$ with respect to the geminal coefficients $\{c_i^a\}$, $\frac{\partial \mathcal{L}}{\partial c_i^a} \vert_{\kappa=0}= 0$, leads to a set of equations for the Lagrange multipliers that are analogous to the pCCD $\Lambda$-equations
\begin{equation}
\label{eq:lagrangian_partial_derivative}
\frac{\partial \mathcal{L}}{\partial c_i^a}  \Big\vert_{\kappa=0} 
    = \langle \Phi_0 | e^{-\Tp} H e^{\Tp}  a_a^\dagger a^\dagger_{\bar a} a_{\bar{i}} a_i | \Phi_0 \rangle 
    + \sum_{j b} \lambda_j^b \langle \Phi_{j{\bar{j}}}^{b{\bar{b}}} |e^{-\Tp} H e^{\Tp} a_a^\dagger a^\dagger_{\bar a} a_{\bar{i}} a_i | \Phi_0 \rangle
\end{equation}
The variational orbital gradient is defined as the partial derivative of the energy with respect to the orbital rotation coefficients $\{\kappa_{pq} \}$,
\begin{align}
\label{eq:variational_orbital_gradient}
\frac{\partial \mathcal{L}}{\partial \kappa_{pq}}  \Big\vert_{\kappa=0} = g_{pq}
&= \langle \Phi_0 | 
e^{-\Tp} 
\big[ (E_{pq} - E_{qp}), H \big] 
e^{\Tp} 
| \Phi_0 \rangle \nonumber \\
&\quad + \sum_{i,a} \lambda^a_i \,
\langle \Phi^{\bar{a}\bar{a}}_{\bar{i}\bar{i}} | 
e^{-\Tp} 
\big[ (E_{pq} - E_{qp}), H \big] 
e^{\Tp} 
| \Phi_0 \rangle \nonumber \\
&= \langle \Phi_0 | (1+\Lambda)
e^{-\Tp} 
\big[ (E_{pq} - E_{qp}), H \big] 
e^{\Tp} 
| \Phi_0 \rangle,
\end{align}
where we introduced the de-excitation operator of pCCD, 
\begin{equation}
    \Lambda = \sum_{ia}  \lambda_i^a a^\dagger_i a^\dagger_{\bar{i}} a_{\bar{a}} a_a 
\end{equation}
The natural orbitals resulting from the orbital optimization within pCCD are typically localized.
By focusing exclusively on electron-pair excitations, the pCCD approach significantly reduces computational cost while improving the description of strongly correlated systems.
This makes it particularly suitable for studying systems with (quasi-)degeneracies,~\cite{diatomics-oo-ap1rog-jpca-2014,ap1rog-singlet-gs-actinides-pccp-2015} such as extended organic molecules.~\cite{pccd-delaram-rsc-adv-2023, ea-eom-pccd-jpca-2024, dip-fpccsd-jctc-2025}
We should stress that the term variational in orbital-optimized pCCD refers solely to the optimization of the orbital rotation parameters.
The optimal molecular orbitals result from minimizing the energy Lagrangian given by Eq.~\eqref{eq:energy_lagrangian_AP1roG}, whereas the geminal coefficients are obtained through the projected Schrödinger equation.


\subsection{N-particle Reduced Density Matrices}
The working equations for the orbital gradient can be efficiently rewritten using 1- and 2-particle reduced density matrices (RDMs).
Since pCCD is a product of natural geminals, the (total) response 1-RDM is diagonal and is calculated from
\begin{equation}\label{eq:1rdm}
    \gamma^p_p = \langle \Phi_0 | (1+\Lambda)
        e^{-\Tp} a^\dagger_p a_p e^{\Tp} 
        | \Phi_0 \rangle
\end{equation}
The response 2-RDM is defined as
\begin{equation}\label{eq:2rdm}
    \Gamma^{pq}_{rs} = \langle \Phi_0 | (1+\Lambda)
        e^{-\Tp} a^\dagger_p a^\dagger_q a_s a_r e^{\Tp} 
        | \Phi_0 \rangle
\end{equation}
Since we work with natural geminals (a seniority-zero wavefunction) and molecular orbitals, the only non-zero blocks of the (spin-integrated) 2-RDM are $\Gamma^{pq}_{pq}$, $\Gamma^{p \bar q}_{p \bar q}$, and $\Gamma^{p\bar p}_{q \bar q}$.
We should note that we do not store the antisymmetrized block of $\Gamma^{pq}_{pq}$, where $\Gamma^{pq}_{pq} = -\Gamma^{pq}_{qp}$, to optimize storage.
Doing so, we further have the relations $\Gamma^{pq}_{pq} = \Gamma^{p \bar q}_{p \bar q}$, $\Gamma^{pq}_{pq} = \Gamma^{qp}_{qp} $, and ${\Gamma}_{p \bar p}^{r \bar r} \neq {\Gamma}_{r \bar r}^{p \bar p}$, while ${\Gamma}_{p \bar p}^{p \bar p} = \gamma^p_p $.
Thus, in the molecular orbital basis, we have to store only 2 blocks of the general 2-RDM $\Gamma^{pq}_{rs}$. Specifically, we have for the total 1- and 2-RDMs
\begin{align}\label{eq:rdms}
    \gamma^{i}_{i} &= 1 - \sum_c c^c_i \lambda_i^c, 
    & \quad 
    \gamma^{a}_{a} &= \sum_k c^a_k \lambda_k^a
\end{align}
\begin{align}\label{eq:rdms_all}
\Gamma^{i\bar{j}}_{i\bar{j}} =\Gamma_{ij}^{ij} &= {}^{\text{corr}}\Gamma^{i\bar{j}}_{i\bar{j}} + {}^{\text{HF}}\Gamma^{i\bar{j}}_{i\bar{j}} + {}^{\text{corr}}\gamma_i^i +{}^{\text{corr}} \gamma_j^j \quad \forall i\neq j\\
%
 &=  1 - \sum_c \lambda_i^c c_i^c - \sum_c \lambda_j^c c_j^c \nonumber \\
\Gamma^{i\bar{a}}_{i\bar{a}} = \Gamma^{a\bar{i}}_{a\bar{i}} = \Gamma^{ia}_{ia} = \Gamma^{ai}_{ai}&={}^{\text{corr}}\Gamma^{i\bar{a}}_{i\bar{a}} + {}^{\text{corr}}\gamma^a_a \\
&= \sum_k \lambda_k^a c_k^a - \lambda_i^a c_i^a \nonumber \\
\Gamma^{i\bar{i}}_{j\bar{j}} &= {}^{\text{corr}}\Gamma^{i\bar{i}}_{i\bar{j}} + 2 {}^{\text{corr}}\gamma_i^i \delta_{ij} + {}^{\text{HF}}\gamma^i_i \delta_{ij} \\
&= \sum_c \lambda_j^c c_i^c + \delta_{ij} (1 - 2 \sum_c \lambda_i^c c_i^c) \nonumber \\
\Gamma^{i\bar{i}}_{a\bar{a}} &= {}^{\text{corr}}\Gamma^{i\bar{i}}_{a\bar{a}}  \\[4pt]
&= c_i^a +2\lambda ^a_i c^a_ic^a_i -2\sum_k \lambda_k^a c_k^a c_i^a - 2 \sum_c \lambda_i^b c_i^c c_i^c + \sum_{kc}\lambda_k^c c_i^c c_k^a\nonumber \\
\Gamma^{a\bar{a}}_{i\bar{i}} &= {}^{\text{corr}}\Gamma^{a\bar{a}}_{i\bar{i}}  \\[4pt]
&= \lambda_i^a\nonumber \\
\Gamma^{a\bar{a}}_{b\bar{b}} &= {}^{\text{corr}}\Gamma^{a\bar{a}}_{b\bar{b}} \\[4pt]
&= \sum_k \lambda_k^a c_k^b
\end{align}
where ${}^{\text{corr}}\gamma$ and ${}^{\text{corr}}\Gamma$ indicate the (pCCD) correlation part of the 1- and 2-RDM, while ${}^{\text{HF}}\gamma$ or ${}^{\text{HF}}\Gamma$ encodes the 1- and 2-RDM of the single Slater determinant reference function.
Exploiting the 1- and 2-RDMs from above, we can rewrite the equations for the orbital gradient of Eq.~\eqref{eq:variational_orbital_gradient} in compact form (again for spatial orbitals)
\begin{align}
\label{eq:variational_orbital_gradient-1}
g_{pq} &= 4h_{pq} \gamma_{q}^q + 4 \sum_{r} \Big(
\langle q r || p r \rangle \Gamma_{qr}^{qr}
+ \langle q \bar{r} || p \bar{r} \rangle \Gamma_{q\bar{r}}^{q\bar{r}}
+ \langle p \bar{q} || r \bar{r} \rangle \tilde{\Gamma}_{q \bar q}^{r \bar r} \Big) \nonumber \\
&- 4h_{qp} \gamma_p^p  - 4 \sum_{r} \Big(
  \langle q r || p r \rangle \Gamma_{pr}^{pr}
+ \langle q \bar{r} || p \bar{r} \rangle \Gamma_{p\bar{r}}^{p\bar{r}}
+ \langle q \bar{p} || r \bar{r} \rangle \tilde{\Gamma}_{p \bar p}^{r \bar r}
\Big) \nonumber \\
&= 4 F_{pq} - 4 F_{qp},
\end{align}
where we have introduced the symmetrized 2-RDM block $\tilde{\Gamma}_{p \bar p}^{r \bar r} = \nicefrac{1}{2} ( {\Gamma}_{p \bar p}^{r \bar r} + {\Gamma}_{r \bar r}^{p \bar p})$, $\langle pq || rs \rangle$ are the two-electron integrals in the physicists notation, and the generalized Fock matrix (GFM) of pCCD
\begin{align}
\label{eq:gfm}
F_{pq} &= h_{pq} \gamma^q_{q} + \sum_{r} \Big(
\langle q r || p r \rangle \Gamma_{qr}^{qr}
+ \langle q \bar{r} || p \bar{r} \rangle \Gamma_{q\bar{r}}^{q\bar{r}}
+ \langle p \bar{q} || r \bar{r} \rangle \tilde{\Gamma}_{q \bar q}^{r \bar r} \Big)
\end{align}

\subsection{Extended Koopmans' Theorem}
Quantum chemical approaches including Hartree--Fock, multi-configuration self-consistent field (MCSCF), and configuration interaction (CI) determine molecular energies through Hamiltonian expectation values. 
The extended Koopmans' theorem (EKT) formalism and its operational equations have been rigorously developed for these expectation-value-based methods, as documented in previous studies~\cite{bozkaya-oo-ccd-jcp-2011,ekt-ip-bozkaya-jcp-2013}.
In contrast, projective methodologies like Møller-Plesset (MP) perturbation theory and coupled-cluster (CC) techniques employ distinct optimization strategies rooted in projection operators.
While Cioslowski \textit{et al.}~\cite{ekt-cioslowski-jcp-1997} demonstrated preliminary EKT implementations for MP2, MP3, and QCISD frameworks, their methodology was based on a mixed approach.
They retained expectation-value formulations for ionized states while using projective calculations for neutral states.
Bozkaya extended the EKT beyond the expectation-value approach for an orbital-optimized CC wavefunction.~\cite{ekt-ip-bozkaya-jcp-2013}
This work presents an EKT derivation adapted for orbital-optimized pCCD wavefunctions.

In the following, we consider electronic states for an $N$-electron system $|\Psi^N\rangle$
\begin{equation}
    |\Psi^N ) = e^{\Tp}|\Phi^N\rangle,
    \label{eqn:cluster-wavefunction}
\end{equation}
which satisfy the Schrödinger equation
\begin{equation}
    H|\Psi^N ) = E^N|\Psi^N )
    \label{eqn:neutral_state}
\end{equation}

The energy eigenvalue of the neutral state is obtained through projection
\begin{equation}
    E^N = (\Phi^N|H|\Psi^N ),
    \label{eqn:energy-neutral_state}
\end{equation}
where
\begin{equation}
    (\Phi^N| = \langle\Phi^N|{(1+\Lambda)} e^{-\Tp},
    \label{eqn:bra}
\end{equation}
and
\begin{equation}
    |\Psi^N ) = e^{\Tp}|\Phi^N\rangle,
    \label{eqn:ket}
\end{equation}
with intermediate normalization
\begin{equation}
\begin{aligned}
( \Phi^N | \Psi^N ) 
&= \langle \Phi^N | (1 + \Lambda) \, e^{-\Tp} e^{\Tp} | \Phi^N \rangle \\
&= \langle \Phi^N | 1 + \Lambda | \Phi^N \rangle \\
&= \langle \Phi^N | \Phi^N \rangle + \langle \Phi^N | \Lambda | \Phi^N \rangle \\
&= 1
\end{aligned}
\label{eq:intermediate-normalization}
\end{equation}
as $\Lambda$ is a de-excitation operator and hence $\Lambda |\Phi^N\rangle = 0$.
For the energy expectation value, we get
\begin{equation}
\begin{aligned}
E^N 
&= ( \Phi^N | \bar{H} | \Psi^N ) \\
&= \langle \Phi^N | (1 + \Lambda) e^{-\Tp} H e^{\Tp} | \Phi^N \rangle \\
&= \langle \Phi^N | \bar{H} | \Phi^N \rangle + \langle \Phi^N | \Lambda \bar{H} | \Phi^N \rangle \\
&= \langle \Phi^N | \bar{H} | \Phi^N \rangle
\end{aligned}
\label{eq:energy-cc}
\end{equation}
as the pCCD amplitude equations are satisfied, $\langle \Phi^N | \Lambda \bar{H} | \Phi^N \rangle =0$.
For the $(N-1)$-electron ionized state, we have the analogous relationship
\begin{equation}
    E^{N-1} = (\Phi^{N-1}|H|\Psi^{N-1}),
    \label{eqn:n-1_state}
\end{equation}
where we express the $(N-1)$-electron states in terms of the $N$-electron system with wavefunctions
\begin{align}
    |\Psi^{N-1} ) &= A|\Psi^N ) =Ae^{\Tp}|\Phi^N\rangle, \\
    ( \Phi^{N-1}| &=  (\Phi^N|A^\dagger= \langle\Phi^N|{(1+\Lambda)} e^{-\Tp}A^\dagger,
    \label{eqn:a-operator}
\end{align}
using the ionization operators $A$ and its adjoint
\begin{align}
    A &= \sum_p c_p a_p, \\
    A^\dagger &= \sum_p c_p^* a_p^\dagger
    \label{eqn:n-1_a},
\end{align}
where $\{c_p\}$ denote the expansion coefficients (to be optimized) and we assumed the removal of $\alpha$ electrons.
Hence, in the following, little letters indicate $\alpha$ spin degrees of freedom.
It is also assumed that the $(N-1)$-electron state fulfills the intermediate normalization condition expressed as
\begin{equation}
    ( \Phi^{N-1} | \Psi^{N-1} ) = \langle \Phi^{N}|{(1+\Lambda)} e^{\Tp} A^\dagger A e^{\Tp}| \Psi^{N} \rangle = 1
    \label{eqn:n-1-cond}
\end{equation}
Now, we define an energy Lagrangian for the ionized state, which ensures that the normalization condition above is satisfied, namely
\begin{align}
    \mathcal{L}^{{\rm{{IP}}}} &= E^{N-1} - E^N + \mathbf{e} \left( ( \Phi^{N-1} |\Psi^{N-1} ) - 1 \right) \nonumber \\
                     &= E^{N-1} - E^N + \mathbf{e} \left( ( \Phi^{N} | A^\dagger A |\Psi^{N} ) - 1 \right)
    \label{eqn:lagrange}
\end{align}
Substituting the energy expressions Eqs.~\eqref{eq:energy-cc} and \eqref{eqn:n-1_state} into the above equation yields
\begin{equation}
    \mathcal{L}^{{\rm{{IP}}}} = ( \Phi^{N} | A^\dagger H A | \Psi^{N} ) - ( \Phi^{N} | H | \Psi^{N} ) + \mathbf{e} \left( ( \Phi^{N} | A^\dagger A | \Psi^{N} ) - 1 \right)
    \label{eqn:lagrange-2}
\end{equation}
Since $|\Psi^{N})$ is an eigenfunction of $H$, $E^N = H|\Psi^{N})$, the second term in Eq.~\eqref{eqn:lagrange-2} can be expressed as
\begin{equation}
    ( \Phi^{N} | H | \Psi^{N} ) = ( \Phi^{N} | A^\dagger A H | \Psi^{N} )= E^{N} ( \Phi^{N} | A^\dagger A| \Psi^{N} )
    \label{eqn:lagrange-3}
\end{equation}
and we can rewrite the Lagrangian of Eq.~\eqref{eqn:lagrange-2} as
\begin{align}
    \mathcal{L}^{{\rm{{IP}}}} 
    &=  ( \Phi^{N} | A^\dagger H A | \Psi^{N} )  - ( \Phi^{N} | A^\dagger A H | \Psi^{N} )+ \mathbf{e} \left( ( \Phi^{N} | A^\dagger A | \Psi^{N} ) - 1 \right), \nonumber \\
    &= ( \Phi^{N} | A^\dagger [H, A] | \Psi^{N} ) + \mathbf{e} \left( ( \Phi^{N} | A^\dagger A | \Psi^{N} ) - 1 \right)
    \label{eqn:lagrange-4}
\end{align}
Substituting the expression for the ionization operator and its adjoint, the above Lagrangian reads
\begin{align}
    \mathcal{L}^{{\rm{{IP}}}} 
    &= \sum_{pq}( \Phi^{N} | a_q^\dagger [H, a_p] | \Psi^{N} ) c_q^* c_p + \mathbf{e} \left( \sum_{pq}( \Phi^{N} | a_q^\dagger a_p | \Psi^{N} ) c_q^* c_p  - 1 \right)
    \label{eqn:lagrange-explicit}
\end{align}
The ionized states $|\Psi^{N-1})$ are obtained by minimizing $\mathcal{L}^{IP}$ with respect to $A^\dagger$, which leads to the condition
\begin{equation}
    \frac{\partial \mathcal{L}^{{\rm{{IP}}}}} {\partial c_q^*} =
    \sum_{p}( \Phi^{N} | a_q^\dagger [H, a_p] | \Psi^{N} ) c_p 
    + \mathbf{e} \sum_{p}( \Phi^{N} | a_q^\dagger a_p | \Psi^{N} ) c_p = 0
\end{equation}
More explicitly, this can be written as
\begin{align}
   \frac{\partial \mathcal{L}^{{\rm{{IP}}}}} {\partial c_q^*}
   &= \sum_p c_p \langle \Phi^{N} |{(1+\Lambda)} e^{-\Tp} a_{q}^\dagger [H, a_p] e^{\Tp}| \Phi^{N} \rangle \nonumber \\
   &\phantom{=} + \mathbf{e} \sum_p c_p  \langle \Phi^{N} | {(1+\Lambda)} e^{-\Tp}a_{q}^\dagger a_p e^{\Tp}| \Phi^{N} \rangle \nonumber \\
    &= - \sum_p c_p F_{qp} + e \sum_p\gamma^q_p c_p = 0
    \label{eqn:commutator-cond}
\end{align}
The second term in Eq.~\eqref{eqn:commutator-cond} corresponds to the response 1-RDM of the $N$-electron system, defined as
\begin{equation}
    \gamma^q_{p} = \langle \Phi^{N} |(1+\Lambda) e^{-\Tp} a_{q}^\dagger a_p e^{\Tp} | \Phi^{N} \rangle,
    \label{eq:opdm}
\end{equation}
while the first term of Eq.~\eqref{eqn:commutator-cond} is equivalent to the GFM, also known as the orbital Lagrangian, of Eq.~\eqref{eq:gfm} (indices without subscripts indicate $\alpha$ electrons, while indices with subscripts [$\sigma$, $\tau$] imply a summation over $\alpha$ and $\beta$ spin degrees of freedom)
\begin{align}
    F_{qp} 
    &= - \langle \Phi^{N} | (1+\Lambda) e^{-\Tp}a_{q}^\dagger [H, a_p]  e^{\Tp}| \Phi^{N} \rangle \nonumber \\
    &= - \sum_{r_\sigma s_\sigma} h_{r_\sigma s_\sigma} \langle \Phi^{N} | (1+\Lambda) e^{-\Tp}a_{q}^\dagger [ a^\dagger_{r_\sigma} a_{s_\sigma}, a_p]  e^{\Tp}| \Phi^{N} \rangle \nonumber \\
    &\phantom{=} - \sum_{r_\sigma s_\tau t_\sigma u_\tau} \frac{1}{2} \langle{r_\sigma s_\tau}|{t_\sigma u_\tau}\rangle \langle \Phi^{N} | (1+\Lambda) e^{-\Tp}a_{q}^\dagger [ a^\dagger_{r_\sigma} a_{s_\tau}^\dagger a_{u_\tau} a_{t_\sigma}, a_p]  e^{\Tp}| \Phi^{N} \rangle \nonumber \\
    &= \sum_{s} h_{ps} \langle \Phi^{N} | (1+\Lambda) e^{-\Tp} a_q^\dagger a_s  e^{\Tp}| \Phi^{N} \rangle \nonumber \\
    &\phantom{=} + \sum_{ts_\tau u_\tau} \bra{ps_\tau}\ket{tu_\tau} \langle \Phi^{N} | (1+\Lambda) e^{-\Tp} a_q^\dagger a_{s_\tau}^\dagger a_{u_\tau} a_t e^{\Tp}| \Phi^{N} \rangle \\
    &= \sum_{s} h_{ps} \gamma^q_s + \sum_{ts_\tau u_\tau} \bra{ps_\tau}\ket{tu_\tau} \Gamma^{qs_\tau}_{tu_\tau} \nonumber \\
    &= h_{pq} \gamma_q^q + \langle pu||qu \rangle \Gamma_{qu}^{qu} 
+ \langle p\bar{u}||q\bar{u} \rangle \Gamma_{q\bar{u}}^{q\bar{u}}  +  \langle p\bar{q}||u\bar{u} \rangle \tilde{\Gamma}_{u\bar{u}}^{q\bar{q}} \nonumber
    \label{eq:gfm-2}
\end{align}
where the commutator ensures that $H$ and $a_p$ are connected and
$h_{ps}$ denotes the one-electron Hamiltonian matrix elements, $\langle ps || tu \rangle$ are the antisymmetrized two-electron integrals (note that $\langle p\bar{q}||u\bar{u} \rangle = \langle p\bar{q}|u\bar{u}\rangle$).
 For pCCD, the non-zero blocks of the 1- and 2-RDM are $\gamma^p_q = \gamma^p_{p} \delta_{pq}$, and  $\Gamma^{pq}_{pq}$, $\Gamma^{p \bar q}_{p \bar q}$, and $\Gamma^{p\bar p}_{q \bar q}$ (or equivalently its symmertrized counterpart $\tilde{\Gamma}_{p\bar{p}}^{q\bar{q}}$), which yields the GFM of Eq.~\eqref{eq:gfm}.
Thus, Eq.~\eqref{eqn:commutator-cond} can be recast into the following eigenvalue problem
\begin{equation}
    \sum_q (F_{qp} - e \gamma^q_p) c_q = 0,
    \label{eq:energy-1}
\end{equation}
or equivalently, in matrix form,
\begin{equation}
\mathbf{F} \mathbf{C} = \boldsymbol{\gamma} \mathbf{C} \mathbf{e},
\label{eq:energy-3}
\end{equation}
where $\mathbf{C}$ is the vector of expansion coefficients and $\mathbf{e}$ is the diagonal matrix of eigenvalues.

It is important to note that Eq.~\eqref{eq:energy-3} closely resembles the self-consistent field (SCF) eigenvalue equation, $\mathbf{F C} = \mathbf{S C} \varepsilon$. In the SCF framework, the occupied block of the GFM corresponds to the occupied block of the Fock matrix, while the occupied block of the 1-RDM corresponds to the occupied block of the overlap matrix. Therefore, only the IPs associated with occupied orbitals have physical significance, even though the entire GFM is diagonalized.

We can now apply a similar approach as used in the SCF method by defining the following intermediates
\begin{equation}
    \mathbf{F}' = \boldsymbol{\gamma}^{-1/2} \mathbf{F} \boldsymbol{\gamma}^{-1/2}, \quad \mathbf{C} = \boldsymbol{\gamma}^{-1/2} \mathbf{C}',
    \label{eq:final}
\end{equation}
which allows us to write the final equation as
\begin{equation}
    \mathbf{F}' \mathbf{C}' = \mathbf{C}' \mathbf{e}
    \label{eq:final-eigen}
\end{equation}
For orbital-optimized methods, the GFM and 1-RDM are, in theory, symmetric and the 1-RDM is positive-definite, making the diagonalization procedure well-conditioned.
However, in practice, the GFM is only approximately symmetric due to the finite convergence threshold applied to the orbital gradient.
In contrast, for pCCD without orbital optimization, the GFM is inherently nonsymmetric, as is the case for standard methods such as MP2, MP3, and CEPA(0). 
To address this, Cioslowski et al.~\cite{ekt-cioslowski-jcp-1997} proposed using relaxed GFMs and density matrices that include the orbital response, which restores symmetry.
However, the relaxed 1-RDM may not be positive-definite, potentially violating $N$-representability.
Although numerical techniques can remove negative eigenvalues, they do not fully correct the 1-RDM's qualitative issues.
Moreover, response contributions may degrade the GFM quality and yield unphysical ionization potentials in challenging systems.
Given these limitations, orbital-optimized methods remain the most reliable choice for EKT studies.
In the case of oo-pCCD, the 1-RDM is positive-definite, while the GFM is (approximately) symmetric (within the convergence threshold defined a priori for the orbital gradient).
However, the 1-RDM might feature very small values (or equivalently occupation numbers), which might lead to numerical instabilities when solving Eq.~\eqref{eq:final-eigen} due to the transformation of Eq.~\eqref{eq:final}.
To avoid numerical instabilities, we introduced a cutoff threshold for the oo-pCCD occupation numbers (that is, the 1-RDM), neglecting all occupation numbers and GFM elements for orbitals falling below the truncation threshold (see also section~\ref{sec:comput-det}).

Finally, we emphasize that pCCD-based EKT is formally distinct from the recently introduced modified Koopmans' theorem (MKT) for pCCD,\mbox{~\cite{pccd-ip-ea-mod-koopmans-jcp-2025}} which extends the standard Koopmans' theorem by incorporating pCCD electron correlation effects. While KT/MKT can be understood as a diagonal approximation of the IP/EA-EOM-pCCD Hamiltonian,\mbox{~\cite{ip-eom-pccd-chem-comm-2021, ea-eom-pccd-jpca-2024}} EKT bears no relation to EOM-pCCD-based methods (or to ADC-type approaches).
The extended Koopmans' theorem is, however, related to orbital gradient theory.\mbox{~\cite{ekt-cioslowski-jcp-1997}}

\begin{figure}
\centering
\includegraphics[scale=0.85]{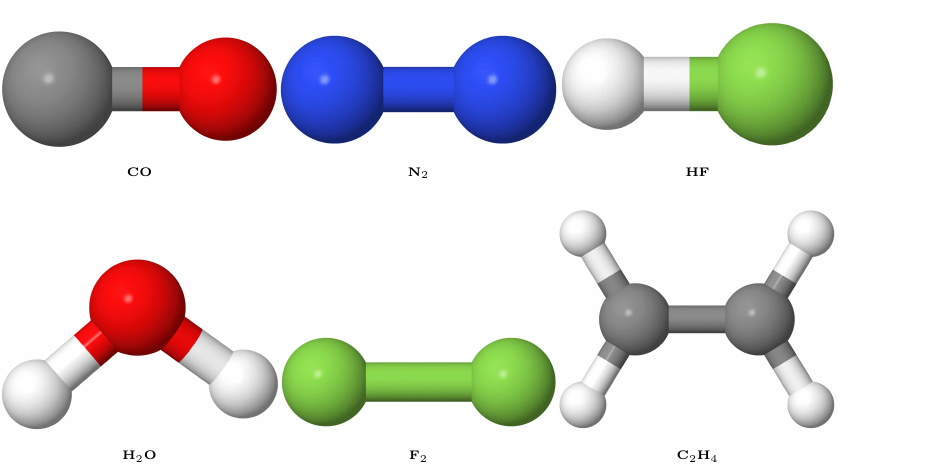}
\caption{The set of six benchmark molecules from Ref.~\citenum{small-molecules-ip-jcp-1996}.}
\label{fig:molecular-structures-2}
\end{figure} 
\begin{figure*}
\centering
\includegraphics[scale=1]{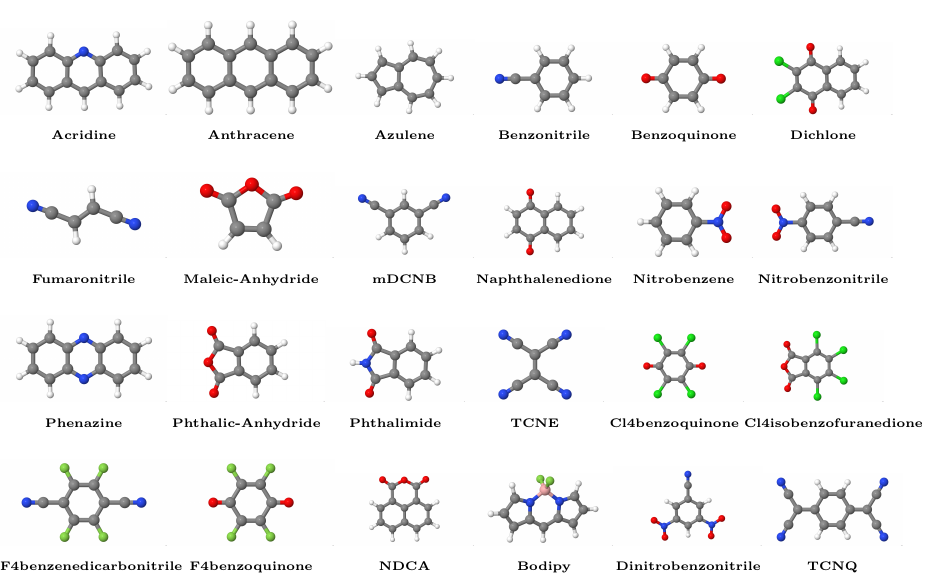}
\caption{The acceptor benchmark data set of 24 organic molecules taken from Refs.~\citenum{24-acceptor-molecules-gw-jctc-2016,richard-24molecules-ip-ea-jctc-2016}.}
\label{fig:molecular-structures-1}
\end{figure*} 

\section{Computational Details}\label{sec:comput-det}
All implementations and calculations were performed using a developer version of the \textsc{PyBEST} software package (v2.2.0dev0).~\cite{pybest-paper-cpc-2021, pybest-paper-cpc-2024} 
We benchmarked ionization potentials for eight neutral atoms (He, Be, Ne, Mg, Ca, Ar, Kr, and Zn) against IP-EOM-pCCD and IP-EOM-fpCCD reference data~\cite{ip-eom-pccd-chem-comm-2021, tailored-ip-eom-pccd-jctc-2024}. 
The correlation-consistent basis set series---cc-pVDZ, cc-pVTZ, and cc-pVQZ---was employed~\cite{dunning-gaussian-basis-set-jcp-1989}.

We analyzed two molecular benchmark data sets. 
The first consists of six small molecules, for which we took the optimized structures from Ref.~\citenum{small-molecules-ip-jcp-1996}. 
Their molecular structures are shown in Figure~\ref{fig:molecular-structures-2} and the corresponding xyz coordinates provided in Tables~S3–S8 of the Supporting Information. 
Experimental IPs for these molecules are available in Ref.~\citenum{kimura-exp-ip-small-molecules-book-1981}, while theoretical data is taken from  Refs.~\citenum{ekt-cioslowski-jcp-1997, chong-Kohn-Sham-ip-jcp-2002, ekt-ip-bozkaya-jcp-2013}. 
The second data set comprises 24 organic acceptor molecules, accompanied by experimental reference data from Ref.~\citenum{richard-24molecules-ip-ea-jctc-2016} and theoretical data from Ref.~\citenum{richard-24molecules-ip-ea-jctc-2016, ip-cc2-adc2-jctc-2022}. 
Their optimized xyz coordinates, obtained from Refs.~\citenum{richard-24molecules-ip-ea-jctc-2016,24-acceptor-molecules-gw-jctc-2016}, are illustrated in Figure~\ref{fig:molecular-structures-1}. 
For both data sets, the cc-pVDZ~\cite{dunning-gaussian-basis-set-jcp-1989}, cc-pVTZ, aug-cc-pVDZ, and aug-cc-pVTZ~\cite{aug-cc-pvtz-jcp-1989} basis sets were employed.

In all calculations, two-electron integrals were approximated using the Cholesky decomposition~\cite{cholesky-review-2011} with a threshold of $10^{-4}$.
All correlated calculations were performed within the frozen-core approximation as implemented in \textsc{PyBEST}, if not stated otherwise.
All computations employed two types of molecular orbitals: 
(i) canonical Hartree–Fock (HF) orbitals and 
(ii) variationally optimized natural pCCD orbitals.~\cite{oo-ap1rog-prb-2014}
Specifically, within our EKT framework, these variants are denoted as EKT(HF) and EKT(pCCD), respectively.
The 1- and 2-RDMs entering the EKT(HF) and EKT(pCCD) equations are defined in the same way for both approaches [Eqs.\mbox{~\ref{eq:1rdm}} and\mbox{~\ref{eq:2rdm}}], with the only difference being the underlying orbitals: HF orbitals in EKT(HF), which yield unrelaxed response density matrices (that is, neglecting the orbital response), and pCCD orbitals in EKT(pCCD), which provide relaxed response density matrices.\mbox{~\cite{oo-ap1rog-prb-2014, pccd-orbital-entanglement-ijqc-2015, pccd-dipole-moments-jctc-2024}}

For atomic EKT calculations, we used different cutoff values for occupation numbers, but for all molecular calculations, a 5$ \times 10^ {-5}$ value was employed. 
All molecular structures were visualized using the Jmol software package.~\cite{jmol}
\section{Results and Discussion}\label{sec:results}
In the following, we evaluate the performance of our newly developed IP computational models for a group of selected closed-shell atoms and two benchmark molecular sets, depicted in Figures~\ref{fig:molecular-structures-2} and~\ref{fig:molecular-structures-1}.
\subsection{Atoms}
We begin our analysis with atoms, for which a highly reliable experimental reference exists.~\cite{nist-2013} 
Specifically, our test set comprises He, Be, Ne, Mg, Ar, Ca, Zn, and Kr atoms, spanning various ranges and types of electron correlation effects.
The noble gases (He, Ne, Ar, and Kr) exhibit weak correlation, while Be, Mg, Ca, and Zn display a mixed pattern of weak and strong electron correlation behavior.
Such diversity makes them ideal candidates for benchmarking new electronic structure methods. 

Table~\ref{tbl:table-ip-atom} compares experimental IPs with Koopmans' and extended Koopmans' values across multiple cutoff values. 
The generalized eigenvalue problem in Eq.~\eqref{eq:final-eigen} is ill-conditioned when the occupation numbers are very small (the oo-pCCD 1-RDM is a diagonal matrix with occupation numbers), so we eliminated natural orbitals with occupation numbers less than a predefined threshold of $5\times 10^{-5}, 5\times 10^{-4}, 1\times 10^{-3}$, and $1\times 10^{-2}$ from our EKT calculations.
That way, we were able to assess the numerical stability and accuracy of the computed IPs. 
It is evident from Table~\ref{tbl:table-ip-atom} that EKT generally improves upon the standard Koopmans' theorem (KT) using canonical HF and natural (localized) pCCD orbitals. 
This improvement stems from the incorporation of electron correlation effects within the pCCD ansatz and its orbital-optimized variant as encoded in the 1-RDM (the 1-RDM does not enter the original KT).
In contrast, the dependence on the cutoff of the occupation numbers (or the 1-RDM) differs substantially between EKT(HF) and EKT(pCCD).
The EKT(HF)-determined IPs for, e.g., Be, Mg, Ca, and Zn exhibit strong variation with the threshold, with larger cutoffs ({$1 \times 10^{-2}$}) consistently yielding better agreement with experiment.
This behavior arises because the canonical HF orbitals yield pCCD 1-RMD with many small numbers, which induce numerical instabilities in the generalized eigenvalue problem of Eq.{~\eqref{eq:final-eigen}}.
The issue is more pronounced for larger basis sets and in the presence of augmented functions (cf. Table~{~\ref{tbl:table-ip-atom}}).
Accordingly, a more aggressive cutoff effectively eliminates these problematic contributions.
These numerical issues with small occupation numbers are significantly mitigated in the EKT(pCCD) approach through the use of natural pCCD orbitals, which are generally known to reduce the basis-set dependence of computed IPs and EAs.{~\cite{ea-eom-pccd-jpca-2024, pccd-ip-ea-mod-koopmans-jcp-2025,ea-eom-fpccd-jctc-2025}}
We emphasize that, in this case, a fully relaxed 1-RDM is obtained from pCCD, rendering the generalized eigenvalue problem of Eq.{~\eqref{eq:final-eigen}} better conditioned---as it was originally formulated for orbital-optimized electron correlation methods.
To this end, excessively large thresholds (e.g., larger than $1\times10^{-3}$ for the atoms listed in Table{~\ref{tbl:Molecules-IP-theoretical}}) begin to discard valuable information when pCCD-optimized orbitals are utilized, thereby worsening the results.
Thus, from now on, we can safely use the cutoff values of $5\times 10^{-5}$ for our EKT calculations once pCCD-optimized orbitals are employed. 
The basis-set convergence of EKT(pCCD) ionization potentials is generally smooth and rapid, with differences between the cc-pVDZ and cc-pVQZ results typically below 0.2 eV.
However, convergence is not strictly monotonic in all cases, most notably for noble gas atoms and zinc.
These systems are dominated by dynamic electron correlation effects, which are particularly challenging to describe accurately within the pCCD framework.
Moreover, the KT basis set dependence is significantly reduced when the pCCD orbitals are utilized. 
The exact match of EKT(pCCD)/cc-pVQZ IP of He with experimental value is to be expected, as our oo-pCCD model is exact for two-electron systems. 

Finally, Table S1 summarizes the statistical performance of our computations with respect to experimental IPs, using mean errors (MEs), mean absolute errors (MAEs), root-mean-square errors (RMSEs), mean percentage errors (MPEs), and standard deviations (SDs). 
Given the mean-field-like computational cost of EKT(pCCD), it achieves remarkable performance, with MPEs below 3\%, both RMSE and SD around 0.5 eV across all atoms and basis set sizes. 
\begin{table}[ht!]
\caption{
Comparison of the first experimental ionization potential---IP$_1$ (in eV) of selected atoms with Koopmans' (KT) and extended Koopmans's (EKT) theorems for HF and pCCD orbitals and various basis set sizes.  
ncore denotes the number of frozen core orbitals when it differs from a default value.
EKT results are reported for different cutoff values of occupation numbers: $5 \times 10^{-5}$, $5\times 10^{-4}$, $1 \times 10^{-3}$, and $1 \times 10^{-2}$.
}
\label{tbl:table-ip-atom}

\resizebox{0.95\textwidth}{!}{
\begin{tabular}{ccc|cc|cccc|cccc}
\multirow{3}{*}{\textbf{\Lower{Atom}}} & \multirow{3}{*}{\textbf{\Lower{Basis set}}} & \multirow{3}{*}{\textbf{\Lower{Exp. [Ref.~\citenum{nist-2013}]}}} 
& \multicolumn{2}{c|}{\textbf{KT [Ref.~\citenum{pccd-ip-ea-mod-koopmans-jcp-2025}]}} & \multicolumn{8}{c}{\textbf{EKT}} \\
\cline{4-5} \cline{6-13}
& & & \Lower{HF} & \Lower{pCCD} & \multicolumn{4}{c}{HF} & \multicolumn{4}{c}{pCCD} \\
\cline{6-9} \cline{10-13}
& & & & & $5\times10^{-5}$ & $5\times10^{-4}$ & $1\times10^{-3}$ & $1\times10^{-2}$ & $5\times 10^{-5}$ & $5\times10^{-4}$ & $1\times10^{-3}$ & $1\times10^{-2}$ \\
\hline
He & cc-pVDZ &  & 24.88 & 24.89 &24.42 & 24.42 & 24.42& 24.42&24.33 &24.33 &24.33 & 25.77  \\
   & cc-pVTZ &  & 24.97 & 24.97 &$-$3.40 & 24.84  & 24.84&24.84 &24.53 & 24.66&24.66 &26.03   \\
   & cc-pVQZ & 24.59 & 24.98 & 24.97 &$-$7.39 &$-$7.39  &24.92 &24.92 &24.59 &24.69 &24.69 & 26.08  \\
&&&&&&&&&&&\\
Be (ncore=0) & cc-pVDZ &  & 8.41 & 8.34 &3.64 & 3.64 &8.95 &9.17 & 9.29&9.29 &9.29 & 9.54  \\
             & cc-pVTZ &  & 8.42 & 8.33 & 2.16& 2.16 & 2.16&9.05 & 9.23&9.31 & 9.31&  9.55 \\
             & cc-pVQZ & 9.32 & 8.42 & 8.34 &$-$1.70 & 8.98 &8.98 &8.98 &9.42 &9.52 &9.54 & 9.54  \\
&&&&&&&&&&&\\
Ne & cc-pVDZ &  & 22.64 & 22.67 &22.29 &22.29  & 22.29&22.29&22.20 & 22.20&22.20 &  23.22  \\
   & cc-pVTZ &  & 23.01 & 23.03 &4.20 & 22.86 & 22.86& 23.46& 22.61& 22.72& 22.72& 23.72  \\
   & cc-pVQZ & 21.56 & 23.10 & 23.13 &$-$5.61 & 22.99 &22.99 & 23.55& 22.75&22.85 &22.85 & 23.87  \\
&&&&&&&&&&&\\ 
Mg (ncore=1) & cc-pVDZ &  & 6.88 & 6.83 & 2.77& 2.77 &2.77 &7.43 & 7.52&7.52 & 7.52& 7.73  \\
             & cc-pVTZ &  & 6.89 & 6.83 &0.44 &1.43  & 1.42&7.34 & 7.48& 7.54& 7.54& 7.75  \\
             & cc-pVQZ & 7.65 & 6.89 & 6.83 & $-$1.66&7.25  &7.25 &7.25 & 7.47&7.54 &7.54 &  7.75 \\
&&&&&&&&&&&\\
Ar & cc-pVDZ &  & 16.00 & 15.96 & 15.93& 15.93 &15.91& 15.91& 15.98& 15.98&15.98 &16.36  \\
   & cc-pVTZ &  & 16.06 & 16.00 & 7.78& 16.12 &16.48 & 16.48& 16.08& 16.16& 16.16& 16.52  \\
   & cc-pVQZ & 15.76 & 16.08 & 16.02&$-$0.24 & 16.45 &16.45&16.45 &16.13 &16.20 & 16.20& 16.56  \\
&&&&&&&&&&&\\
Ca (ncore=5) & cc-pVDZ &  & 5.32 & 5.28 & 2.15&2.15  &2.15 &5.80 &5.89 & 5.89&5.89 & 6.04  \\
             & cc-pVTZ &  & 5.32 & 5.28 &2.15 & 2.15 & 2.15& 5.80& 5.89& 5.89& 5.89& 6.04 \\
             & cc-pVQZ & 6.11 & 5.12 & 5.28 & $-$6.22& 5.71 &5.71 &5.71& 5.86& 5.90&5.90& 6.05  \\
&&&&&&&&&&&\\
Zn & cc-pVDZ &  & 7.96 & 7.92 &0.46 & 1.93 &8.45 &8.45 & 8.60&8.63 &8.63 & 8.84  \\
   & cc-pVTZ &  & 7.96 & 7.92 &$-$4.56 & 8.44 & 8.44&8.44 & 8.60& 8.62& 8.62& 8.83  \\
   & cc-pVQZ & 9.39 & 7.96 & 7.92 & $-$4.71&8.39  &8.39 & 8.39& 8.63&8.63 & 8.63& 8.84  \\
&&&&&&&&&&&\\
Kr & cc-pVDZ &  & 14.17 & 14.15 & 14.17& 14.17 & 14.46&14.46 & 14.22&14.22 &14.22& 14.49  \\
   & cc-pVTZ &  & 14.25 & 14.22 & 9.57& 14.36 & 14.57& 14.57&14.32 & 14.38&14.38& 14.65  \\
   & cc-pVQZ & 14.00 & 14.26 & 14.22 & $-$1.02& 14.53 & 14.53&14.54 &14.40 &14.41 &14.41& 14.68  \\
\end{tabular}
} 
\end{table}

\subsection{Small Molecules}
Our next test set includes a small group of molecules depicted in Figure~\ref{fig:molecular-structures-2}, for which experimental low-lying IPs are known and which are commonly used as a testing ground for new theoretical models.~\cite{ekt-ip-bozkaya-jcp-2013}    
Our KT(HF/pCCD) and EKT(HF/pCCD) results for cc-pVDZ, cc-pVTZ, aug-cc-pVDZ, and aug-cc-pVTZ basis sets are collected in Table~\ref{tbl:table-ip-molecule}. 
Note that EKT(HF) results for cc-pVDZ and cc-pVTZ basis sets are not shown in Table~\ref{tbl:table-ip-molecule} due to numerical issues (strong cutoff parameter dependence). 
Or data in Table~\ref{tbl:table-ip-molecule} shows that the KT and EKT tend to overestimate IPs relative to experiment.
Deviations from reference data depend on the type of IP (IP$_1$, IP$_2$, IP$_3$, etc.), the type of molecule, and the utilized orbitals (HF vs. pCCD).
EKT is theoretically exact for the lowest-lying IP,~\cite{ekt-exact-ip-jcp-2009} making it less suitable for higher-lying ones.
Specifically, the EKT formalism inadequately accounts for orbital relaxation effects (which are usually more pronounced in larger molecules) and struggles to reproduce the correct long-range decay of the density matrix when computing higher-lying IPs. Accordingly, we observe the largest discrepancies between EKT(pCCD) and experiment for {IP$_3$--IP$_6$} of $\ce{C2H4}$ in Table~\ref{tbl:Molecules-IP-EXPERIMENT}.

In addition, KT(HF) yields IPs that approach experimental values reasonably well, mainly when larger basis sets are employed.
Use of pCCD orbitals worsens the performance of KT. 
The situation is reversed for the EKT approach---much better results are obtained within the pCCD orbital basis. 
While EKT(HF) tends to reduce the overestimation found in KT(HF) results, it significantly underestimates IPs for specific orbitals (e.g., the $(\sigma)$ orbital of $\ce{CO}$).
The EKT(pCCD) approach yields the most reliable IPs and the smallest MAEs, RMSEs, and MPEs across all investigated molecules and basis sets compared to experimental values. 
For specific molecules such as $\ce{CO}$, $\ce{N_2}$, $\ce{HF}$, and $\ce{H_2O}$, EKT(pCCD) consistently provides IPs closest to experimental references. 
The statistical analysis of the EKT(pCCD) approach confirms its marginal dependence on the basis set size.
Specifically, the EKT(pCCD) IPs using the cc-pVDZ basis set are very similar to those with the aug-cc-pVTZ basis set. 
We also note that the accuracy between EKT(pCCD) and reference IPs, similar to other theoretical methods, deteriorates for higher-lying IPs (see, for example, the results for the \ce{C2H4} molecule in Table~\ref{tbl:table-ip-molecule}).
Overall, the newly introduced EKT(pCCD) approach yields highly reliable results for the investigated molecular test set, demonstrating negligible dependence on the basis set size.
\begin{table}[ht!]
\caption{
Comparison of experimental and computed low-lying IPs (in eV) for the molecule test set from Figure~\ref{fig:molecular-structures-2}. 
Calculations were performed using KT and EKT approaches with different orbital bases (HF and pCCD) and various basis set sizes. 
The basis sets cc-pVDZ, cc-pVTZ, aug-cc-pVDZ, and aug-cc-pVTZ are abbreviated as DZ, TZ, aug-DZ, and aug-TZ, respectively. 
Statistical error metrics (in eV and $\%$)---including the mean error (ME), mean absolute error (MAE), root-mean-square error (RMSE), mean percentage error (MPE), and standard deviation (SD)---are reported for each method and basis set combination.
}

\label{tbl:table-ip-molecule}

\resizebox{1\textwidth}{!}{%
\begin{tabular}{lcccccc|cccc|cc|cccc}
\hline
\multirow{3}{*}{\textbf{Molecule}} & 
\multirow{3}{*}{\textbf{Orbital}} & 
\multirow{3}{*}{\textbf{Exp.[Ref.~\citenum{kimura-exp-ip-small-molecules-book-1981}]}} &  
\multicolumn{8}{c|}{\textbf{KT}} & 
\multicolumn{6}{c}{\textbf{EKT}} \\
\cline{4-11} \cline{12-17}
& & & \multicolumn{4}{c|}{HF} & \multicolumn{4}{c|}{pCCD} & \multicolumn{2}{c|}{HF} & \multicolumn{4}{c}{pCCD} \\
\cline{4-7} \cline{8-11} \cline{12-13} \cline{14-17}
& & & DZ & TZ & aug-DZ & aug-TZ & DZ & TZ & aug-DZ & aug-TZ & aug-DZ & aug-TZ & DZ & TZ & aug-DZ & aug-TZ \\
\hline
\ce{CO} & $\sigma$ (IP$_1$)  & 14.01 & 14.96 & 15.09 & 15.12 & 15.13 & 16.00 & 15.97 & 16.01 & 15.99 & 15.40 & 15.40 & 14.41 & 14.57 & 14.56 & 14.62 \\
   & $\pi$   (IP$_2$)  & 16.85 & 17.16 & 17.28 & 17.34 & 17.32 & 17.23 & 17.33 & 17.43 & 17.39 & 17.55 & 17.65 & 16.94 & 16.93 & 17.14 & 16.95 \\
   & $\sigma$  (IP$_3$)& 19.78 & 21.81 & 21.85 & 21.97 & 21.88 & 25.78 & 25.76 & 25.71 & 25.76 & 22.23 & 22.15 & 21.76 & 21.64 & 21.85 & 21.60 \\
  &&&&&&&&&&&&\\ 
\ce{N}$_2$ & $\sigma_g$ (IP$_1$) & 15.60 & 16.37 & 16.47 & 16.56 & 16.54 & 16.26 & 16.39 & 16.48 & 16.47 & 17.04 & 16.98 & 16.37 & 16.57 & 16.88 & 16.65 \\
      & $\pi_u$ (IP$_2$)  & 16.68 & 17.00 & 17.17 & 17.23 & 17.23 & 21.04 & 20.80 & 21.24 & 20.81 & 17.56 & 17.57 & 17.00 & 17.12 & 17.24 & 17.17 \\
      & $\sigma_u$ (IP$_3$)& 18.78 & 21.21 & 21.30 & 21.40 & 21.36 & 25.81 & 20.80 & 25.67 & 20.83 & 21.69 & 21.64 & 21.21 & 20.66 & 21.23 & 20.60 \\
  &&&&&&&&&&&&\\
\ce{HF} & $\pi$ (IP$_1$)   & 16.19 & 17.11 & 17.50 & 17.71 & 17.70 & 17.19 & 17.57 & 17.83 & 17.78 & 18.12 & 18.07 & 16.74 & 17.08 & 17.33 & 17.29 \\
   & $\sigma$ (IP$_2$)& 19.90 & 20.32 & 20.68 & 20.98 & 20.91 & 25.16 & 27.27 & 27.20 & 28.30 & 21.31 & 21.23 & 20.54 & 20.90 & 21.25 & 21.23 \\
   &&&&&&&&&&&&\\
\ce{F}$_2$ & $\pi_g$ (IP$_1$)  & 15.87 & 18.00 & 18.05 & 18.19 & 18.13 & 17.90 & 19.98 & 20.17 & 20.08 & 18.50 & 18.46 & 17.66 & 17.28 & 17.39 & 17.37 \\
      & $\pi_u$ (IP$_2$)  & 18.80 & 20.31 & 20.47 & 20.67 & 20.55 & 20.93 & 19.98 & 20.17 & 20.08 & 21.52 & 21.26 & 21.41 & 21.60 & 21.70 & 21.69 \\
      & $\sigma_g$ (IP$_3$) & 21.10 & 21.99 & 22.05 & 22.19 & 22.13 & 21.88 & 21.28 & 21.55 & 21.42 & 22.47 & 22.43 & 22.19 & 22.40 & 22.69 & 22.52 \\
      &&&&&&&&&&&&\\
\ce{H}$_2$\ce{O} & $b_1$ (IP$_1$) & 12.78 & 13.43 & 13.73 & 13.86 & 13.89 & 13.49 & 13.78 & 13.95 & 13.95 & 14.23 & 14.13 & 13.12 & 13.36 & 13.40 & 13.44 \\
       & $a_1$ (IP$_2$)  & 14.83 & 15.46 & 15.76 & 15.98 & 15.96 & 22.98 & 23.02 & 24.08 & 22.90 & 16.27 & 16.18 & 15.35 & 15.64 & 15.99 & 15.85 \\
       & $b_2$ (IP$_3$) & 18.72 & 18.93 & 19.21 & 19.47 & 19.43 & 22.99 & 24.33 & 24.08 & 24.74 & 19.74 & 19.65 & 19.17 & 19.39 & 19.77 & 19.64 \\
       &&&&&&&&&&&&\\
\ce{C}$_2$\ce{H}$_4$ & $b_{3u}$ (IP$_1$) & 10.68 & 10.18 & 10.24 & 10.25 & 10.27 & 10.18 & 10.26 & 10.31 & 10.28 & 10.58 & 10.56 & 10.79 & 10.79 & 10.79 & 10.76 \\
           & $b_{3g}$ (IP$_2$)  & 12.80 & 13.70 & 13.78 & 13.79 & 13.81 & 15.50 & 18.38 & 18.36 & 18.41 & 13.96 & 13.96 & 13.73 & 13.86 & 13.81 & 13.89 \\
           & $a_g$ (IP$_3$)   & 14.80 & 15.85 & 15.95 & 15.98 & 15.99 & 15.50 & 18.38 & 18.36 & 18.41 & 16.17 & 16.17 & 15.70 & 15.83 & 15.51 & 15.92 \\
           & $b_{2u}$ (IP$_4$) & 16.00 & 17.40 & 17.48 & 17.52 & 17.52 & 21.17 & 18.38 & 18.36 & 18.41 & 17.68 & 17.67 & 17.38 & 15.95 & 15.89 & 15.93 \\
           & $b_{1u}$ (IP$_5$) & 19.10 & 21.47 & 21.52 & 21.58 & 21.57 & 21.17 & 18.38 & 18.36 & 18.41 & 21.72 & 21.70 & 18.35 & 17.49 & 17.49 & 17.52 \\
           & $a_g$ (IP$_6$)  & 23.60 & 28.08 & 28.10 & 28.20 & 28.15 & 22.88 & 22.99 & 23.45 & 23.05 & 28.34 & 28.29 & 20.88 & 21.34 & 18.80 & 21.35 \\
           &&&&&&&&&&&&\\
ME   &   &  & 1.19 & 1.34 & 1.46 & 1.43 & 2.71 & 2.71 & 3.10 &2.83 & 1.76 & 1.71 & 0.69 & 0.68 & 0.69 & 0.76 \\
MAE  &   &  & 1.24 & 1.38 & 1.50 & 1.47 & 2.83 & 2.88 & 3.22 & 2.99 & 1.77 & 1.73 & 1.04 & 1.07 & 1.34 & 1.15 \\
RMSE &   &  & 1.60 & 1.68 & 1.78 & 1.75 & 3.69 & 3.78 & 4.19 & 3.93 & 2.03 & 2.00 & 1.32 & 1.29 & 1.72 & 1.35 \\
MPE  &   &  & 7.08 & 7.93 & 8.59 & 8.44 & 16.75 & 17.61 & 19.55 & 18.22 & 10.15 & 9.90 & 5.86 & 6.08 & 7.43 & 6.53 \\
SD   &   &  & 1.90 & 1.04 & 1.04 & 1.03 & 2.56 & 2.71 & 2.89 & 2.79 & 1.03 & 1.01 & 1.15 & 1.22 & 1.61 & 1.14 \\
\end{tabular}
} 
\end{table}

\subsection{Acceptor Molecules}
Accurate and reliable prediction of ionization potentials (IPs) is paramount for understanding charge transfer processes in organic electronic materials, enabling the enhancement of device performance. 
Notably, in extended organic systems, dynamic electron correlation effects are less pronounced, rendering electron-pair theories, such as pair-coupled cluster doubles (pCCD), effective and reliable.~\cite{pccd-delaram-rsc-adv-2023, pccd-perspective-jpcl-2023, ea-eom-pccd-jpca-2024, ea-eom-fpccd-jctc-2025, pccd-ip-ea-mod-koopmans-jcp-2025} 
This synergy of accuracy and efficiency has been repeatedly validated, underscoring pCCD's reliability for capturing essential correlation without prohibitive costs.
A compelling illustration is the benchmark organic acceptor dataset~\cite{24-acceptor-molecules-gw-jctc-2016}, which features 24 diverse organic molecules with robust, experimentally corroborated reference data alongside high-level theoretical benchmarks—providing an ideal testing ground for method validation.

\begin{figure}
\centering
\includegraphics[scale=0.95]{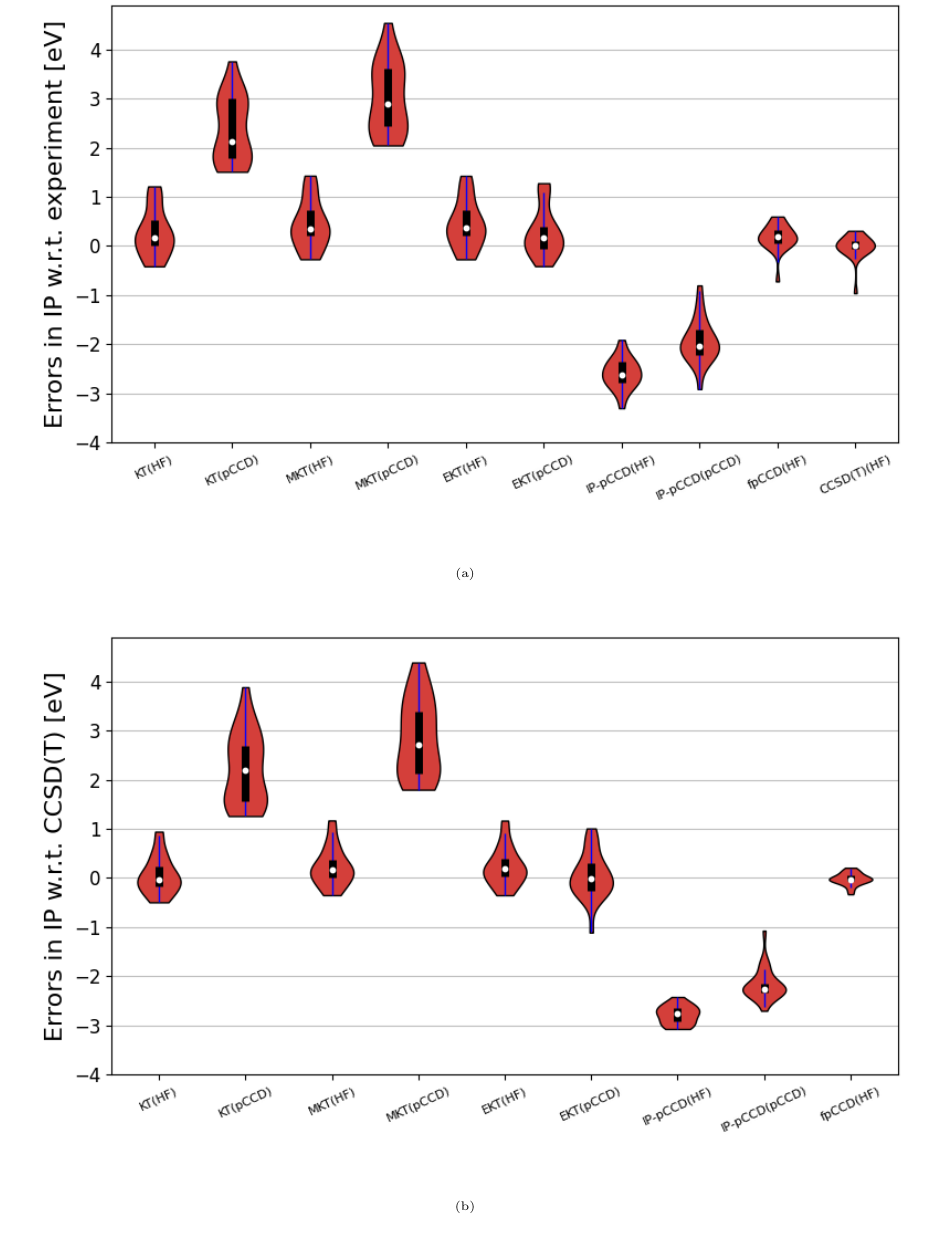}
\caption{Errors in IPs (eV) relative to experimental (a) and theoretical (b) values using the aug-cc-pVDZ basis set.
Note that, due to missing experimental data, IPs were calculated for 21 molecules (see also Table~S2 of the SI).
}
\label{fig:errors-ip-molecules_exp}
\end{figure}

The performance of our EKT(HF/pCCD) theoretical models for predicting IPs in the acceptor data set (shown in Figure~\ref{fig:molecular-structures-1}) is summarized in Figure~\ref{fig:errors-ip-molecules_exp} in terms of violin plots; all individual IPs are provided in Table S2 of the Supporting Information. 
The violin plot is a statistical visualization that combines elements of a box plot and a kernel density estimation to display the distribution of a dataset in a compact and informative way, as shown here for IPs from different methods. 
Specifically, wider sections indicate higher data density (more observations clustering around those values), while narrower sections show sparser data.
This reveals the whole shape of the distribution, including multimodality (multiple peaks) or skewness.
The upper part of Figure~\ref{fig:errors-ip-molecules_exp} compares the EKT(HF/pCCD) IPs and other theoretical models based on pCCD and the CCSD(T) approach (using the standard canonical HF orbitals) to experimental values. 

Among the established pCCD-based approaches, we distinguish between the original KT and its modified variant (MKT),~\cite{pccd-ip-ea-mod-koopmans-jcp-2025} which appixomate IPs through orbital energies, the simple IP-EOM-pCCD model,~\cite{ip-eom-pccd-chem-comm-2021} and its frozen-pair variant—IP-EOM-fpCCD---which includes a significant amount of dynamical correlation on top of pCCD.~\cite{tailored-ip-eom-pccd-jctc-2024}
The latter can be regarded as the most accurate of the investigated pCCD-based models, albeit also the most computationally demanding (with a computational scaling similar to IP-EOM-CCSD). 

Figure~\ref{fig:errors-ip-molecules_exp}(a) demonstrates that CCSD(T) delivers the most accurate IP predictions compared to experiment. 
It is also evident from Figure~\ref{fig:errors-ip-molecules_exp}(a) that EKT(HF) and EKT(pCCD) improve upon the KT(pCCD), MKT(pCCD), and simple IP-EOM-pCCD(HF/pCCD) methods. 
The performance of EKT(pCCD) is better than that of EKT(HF). 
The EKT(pCCD) results have an overall smaller spread, and the data is centered more towards the middle of the experimental reference data. 
The performance of EKT(pCCD) is approaching that of the IP-EOM-fpCCD(HF) method. 
The bottom part of Figure~\ref{fig:errors-ip-molecules_exp} assesses the performance of the pCCD-based methods with respect to CCSD(T) using the aug-cc-pVDZ basis set. 
Here, we also see that the EKT(pCCD) is mainly centered around the reference point, similar to IP-EOM-fpCCD(HF); however, the error spread is larger in EKT(pCCD).

\begin{figure}
\centering
\includegraphics[scale=0.95]{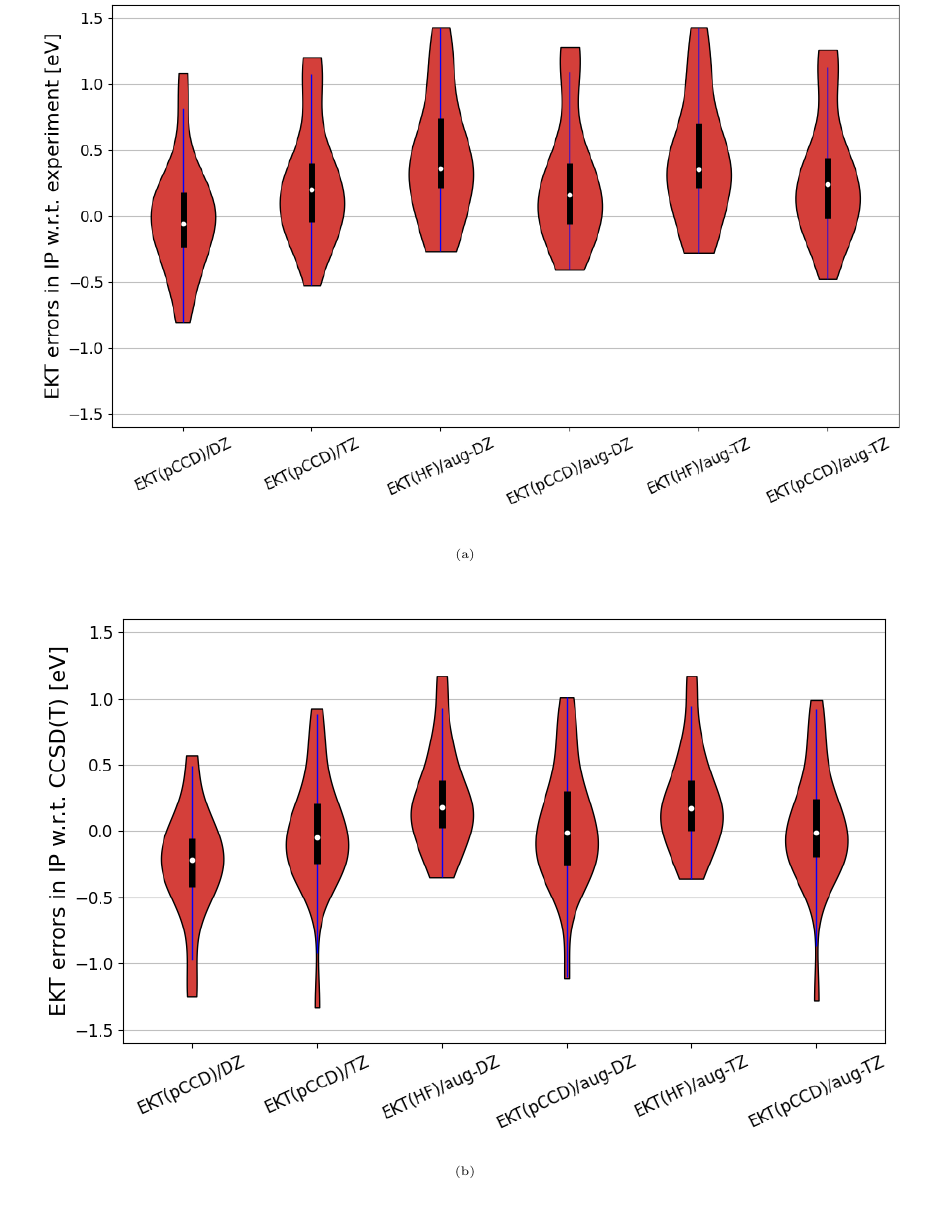}
\caption{EKT(HF/pCCD) errors in IP for various basis set sizes relative to (a) experimental and (b) CCSD(T) data. 
The basis sets cc-pVDZ, cc-pVTZ, aug-cc-pVDZ, and aug-cc-pVTZ are abbreviated as DZ, TZ, aug-DZ, and aug-TZ, respectively. 
Note that, due to incomplete experimental data, IPs were calculated for 21 molecules (see also Table~S2 of the SI).
}
\label{fig:errors-ip-molecules-all-basis}
\end{figure}

Statistical performance with respect to experiment and CCSD(T) using different basis sets is summarized in Tables~\ref{tbl:Molecules-IP-EXPERIMENT} and~\ref{tbl:Molecules-IP-theoretical}, respectively.
From Table~\ref{tbl:Molecules-IP-EXPERIMENT} we see that both EKT(HF) and EKT(pCCD) improve upon the previously developed MKT~\cite{pccd-ip-ea-mod-koopmans-jcp-2025} and simple IP-EOM-pCCD type models.~\cite{ip-eom-pccd-chem-comm-2021}
Note that for the EKT(HF) calculations with small basis sets (cc-pVDZ and cc-pVTZ), we encountered numerical issues, thus the results are not showed in Table~\ref{tbl:Molecules-IP-EXPERIMENT}.
The EKT(pCCD) MAE and MPE errors are smaller than for the EKT(HF) approach. 
Notably, EKT(pCCD) works well across all the basis set sizes with a similar MPE and SD across all basis sets sizes. 
The basis set dependence of EKT(HF) and EKT(pCCD) methods is depicted in Figure~\ref{fig:errors-ip-molecules-all-basis} w.r.t. experiment and CCSD(T). 
In both cases the EKT(pCCD) has a smaller basis set dependence that the EKT(HF) approach.
Addition of augmented functions does not change the overall EKT(pCCD) performance. 
Such an observation is consistent with our previous findings on the modified Koopmans' theorem approach employing pCCD orbitals,~\cite{pccd-ip-ea-mod-koopmans-jcp-2025} as well as with results obtained using EA-EOM-pCCD-based methods.~\cite{ea-eom-pccd-jpca-2024, ea-eom-fpccd-jctc-2025}
This behavior can be attributed to the localized nature of the pCCD-optimized orbitals, which provide a more compact (less diffuse) representation of the wave function compared to the canonical HF orbital basis.

In summary, EKT(pCCD) provides reliable IPs for the acceptor data set with accuracy approaching that of IP-EOM-fpCCD. 
Moreover, the EKT(pCCD) basis set size dependence is negligible, allowing for reliable IP predictions even with small basis sets, such as the cc-pVDZ basis set.
\begin{table}[ht!]
\caption{
Statistical errors with respect to experimental IPs from Refs.~\citenum{richard-24molecules-ip-ea-jctc-2016}, 
computed for 21 acceptor molecules using four different basis sets. 
Error metrics include Mean Error (ME), Mean Absolute Error (MAE), Root-Mean-Square Error (RMSE), 
}
\label{tbl:Molecules-IP-EXPERIMENT}

\begingroup
\footnotesize
\setlength{\tabcolsep}{4pt} 
\renewcommand{\arraystretch}{1} 

\resizebox{\textwidth}{!}{%
\begin{tabular}{llrrrrr}
\hline
\multicolumn{1}{c}{\textbf{Method}} &
\multicolumn{1}{c}{\textbf{Basis set}} &
\multicolumn{1}{c}{\textbf{ME [eV]}} &
\multicolumn{1}{c}{\textbf{MAE [eV]}} &
\multicolumn{1}{c}{\textbf{RMSE [eV]}} &
\multicolumn{1}{c}{\textbf{MPE [\%]}} &
\multicolumn{1}{c}{\textbf{SD [eV]}} \\
\hline
KT(HF) [Ref.~\citenum{pccd-ip-ea-mod-koopmans-jcp-2025}]& cc-pVDZ     &  0.20 & 0.39 & 0.52 &  3.99 & 0.49 \\
              & cc-pVTZ     &  0.24 & 0.40 & 0.53 &  4.09 & 0.49 \\
              & aug-cc-pVDZ &  0.27 & 0.41 & 0.55 &  4.21 & 0.49 \\
              & aug-cc-pVTZ &  0.27 & 0.42 & 0.55 &  4.22 & 0.49 \\
\hline
KT(pCCD) [Ref.~\citenum{pccd-ip-ea-mod-koopmans-jcp-2025}] & cc-pVDZ     &  2.27 & 2.27 & 2.37 & 23.86 & 0.68 \\
                & aug-cc-pVDZ &  2.36 & 2.36 & 2.45 & 24.80 & 0.67 \\
                & aug-cc-pVTZ &  2.31 & 2.31 & 2.39 & 24.20 & 0.65 \\
\hline
MKT(HF) [Ref.~\citenum{pccd-ip-ea-mod-koopmans-jcp-2025}] & cc-pVDZ & 0.43 & 0.51 & 0.65 &5.10 & 0.50 \\
& aug-cc-pVDZ & 0.44 & 0.51 & 0.66 & 5.06 & 0.50 \\
& aug-cc-pVTZ & 0.42 & 0.50 & 0.64 & 4.96 & 0.49 \\
\hline
MKT(pCCD) [Ref.~\citenum{pccd-ip-ea-mod-koopmans-jcp-2025}] & cc-pVDZ & 2.90 & 2.90 & 2.99 & 30.37 & 0.73 \\
& aug-cc-pVDZ & 3.01 & 3.01 & 3.09 & 31.43 & 0.73 \\
& aug-cc-pVTZ & 2.94 & 2.94 & 3.02 & 30.76 & 0.72 \\
\hline
IP-EOM-pCCD(HF) [Ref.~\citenum{tailored-ip-eom-pccd-jctc-2024}] & cc-pVDZ     & $-$2.56 & 2.56 & 2.58 & 26.75 & 0.33 \\
                         & aug-cc-pVDZ & $-$2.59 & 2.59 & 2.61 & 27.06 & 0.32 \\
\hline
IP-EOM-pCCD(pCCD) [Ref.~\citenum{tailored-ip-eom-pccd-jctc-2024}] & cc-pVDZ     & $-$1.97 & 1.97 & 2.02 & 20.65 & 0.44 \\
                           & aug-cc-pVDZ & $-$1.94 & 1.94 & 1.99 & 20.32 & 0.45 \\
                           & aug-cc-pVTZ & $-$2.11 & 2.11 & 2.14 & 22.04 & 0.36 \\
\hline
IP-EOM-fpCCD(HF) [Ref.~\citenum{tailored-ip-eom-pccd-jctc-2024}] & aug-cc-pVDZ &  0.18 & 0.25 & 0.32 &  2.51 & 0.28 \\
\hline
EKT(HF) 
                       & aug-cc-pVDZ &   0.44 & 0.51 & 0.65 &  5.08 & 0.49 \\
                       & aug-cc-pVTZ &   0.43 & 0.51 & 0.64 &  5.00 & 0.49 \\
\hline
EKT(pCCD) & cc-pVDZ     & $-$0.01 & 0.32 & 0.43 &  3.18 & 0.44 \\
                         & cc-pVTZ     &   0.26 & 0.38 & 0.53 &  3.73 & 0.48 \\
                         & aug-cc-pVDZ &   0.28 & 0.39 & 0.57 &  3.83 & 0.50 \\
                         & aug-cc-pVTZ &   0.31 & 0.40 & 0.56 &  3.93 & 0.48 \\
\hline
CCSD(T)(HF) [Ref.~\citenum{richard-24molecules-ip-ea-jctc-2016}] & aug-cc-pVDZ & $-$0.01 & 0.15 & 0.25 &  1.46 & 0.26 \\
\hline\hline
\end{tabular}
}
\endgroup
\end{table}

\begin{table}[ht!]
\caption{
Statistical errors with respect to theoretical IPs from Refs.~\citenum{richard-24molecules-ip-ea-jctc-2016, ip-cc2-adc2-jctc-2022}, 
computed for 24 acceptor molecules using four different basis sets. 
Error metrics include Mean Error (ME), Mean Absolute Error (MAE), Root-Mean-Square Error (RMSE), 
Mean Percentage Error (MPE), and Standard Deviation (SD).
}
\label{tbl:Molecules-IP-theoretical}

\begingroup
\footnotesize
\setlength{\tabcolsep}{4pt} 
\renewcommand{\arraystretch}{1} 

\resizebox{\textwidth}{!}{%
\begin{tabular}{llrrrrr}
\hline
\multicolumn{1}{c}{\textbf{Method}} &
\multicolumn{1}{c}{\textbf{Basis set}} &
\multicolumn{1}{c}{\textbf{ME [eV]}} &
\multicolumn{1}{c}{\textbf{MAE [eV]}} &
\multicolumn{1}{c}{\textbf{RMSE [eV]}} &
\multicolumn{1}{c}{\textbf{MPE [\%]}} &
\multicolumn{1}{c}{\textbf{SD [eV]}} \\
\hline
KT(HF) [Ref.~\citenum{pccd-ip-ea-mod-koopmans-jcp-2025}]& cc-pVDZ     & $-$0.02 & 0.31 & 0.38 & 3.29 & 0.41 \\
              & cc-pVTZ     &  0.03 & 0.29 & 0.38 & 3.03 & 0.40 \\
              & aug-cc-pVDZ &  0.06 & 0.29 & 0.38 & 3.00 & 0.41 \\
              & aug-cc-pVTZ &  0.06 & 0.28 & 0.38 & 2.94 & 0.40 \\
\hline
KT(pCCD) [Ref.~\citenum{pccd-ip-ea-mod-koopmans-jcp-2025}] & cc-pVDZ     &  2.11 & 2.11 & 2.24 & 21.96 & 0.77 \\
                & aug-cc-pVDZ &  2.21 & 2.21 & 2.33 & 22.94 & 0.76 \\
                & aug-cc-pVTZ &  2.16 & 2.16 & 2.26 & 22.36 & 0.70 \\
\hline
MKT(HF) [Ref.~\citenum{pccd-ip-ea-mod-koopmans-jcp-2025}] & cc-pVDZ &0.22&0.33& 0.45& 3.25 & 0.42 \\
& aug-cc-pVDZ & 0.23 & 0.33& 0.45 & 3.24 & 0.42 \\
& aug-cc-pVTZ & 0.22 & 0.32& 0.44 & 3.15 & 0.41 \\
\hline
MKT(pCCD) [Ref.~\citenum{pccd-ip-ea-mod-koopmans-jcp-2025}] & cc-pVDZ & 2.74 & 2.74 & 2.85 & 28.33 &0.80 \\
& aug-cc-pVDZ & 2.85 & 2.85 & 2.95 & 29.38 & 0.80\\
& aug-cc-pVTZ & 2.78 & 2.78&  2.88&  28.74& 0.76 \\
\hline
IP-EOM-pCCD(HF) [Ref.~\citenum{tailored-ip-eom-pccd-jctc-2024}] & cc-pVDZ     & $-$2.75 & 2.75 & 2.76 & 28.30 & 0.24 \\
                         & aug-cc-pVDZ & $-$2.78 & 2.78 & 2.79 & 28.62 & 0.20 \\
\hline
IP-EOM-pCCD(pCCD) [Ref.~\citenum{tailored-ip-eom-pccd-jctc-2024}] & cc-pVDZ     & $-$2.20 & 2.20 & 2.22 & 22.68 & 0.31 \\
                           & aug-cc-pVDZ & $-$2.16 & 2.16 & 2.18 & 22.33 & 0.32 \\
                           & aug-cc-pVTZ & $-$2.32 & 2.32 & 2.33 & 23.92 & 0.21 \\
\hline
IP-EOM-fpCCD(HF) [Ref.~\citenum{tailored-ip-eom-pccd-jctc-2024}] & aug-cc-pVDZ & $-$0.02 & 0.09 & 0.12 & 0.95 & 0.10 \\
\hline
EKT(HF) 
                       & aug-cc-pVDZ &   0.23 & 0.33 & 0.45 &  3.25 & 0.41 \\
                       & aug-cc-pVTZ &   0.22 & 0.32 & 0.44 &  3.15 & 0.41 \\
\hline
EKT(pCCD) & cc-pVDZ     & $-$0.24 & 0.37 & 0.49 &  3.75 & 0.41 \\
                         & cc-pVTZ     &   0.00 & 0.32 & 0.44 &  3.23 & 0.37 \\
                         & aug-cc-pVDZ &   0.05 & 0.34 & 0.46 &  3.42 & 0.42 \\
                         & aug-cc-pVTZ &   0.04 & 0.32 & 0.45 &  3.26 & 0.38 \\
\hline\hline
\end{tabular}
}
\endgroup
\end{table}

\section{Conclusions and Outlook}\label{sec:conclusions}
In this work, we derived the working equations and implemented the EKT on top of the pCCD wavefunction.
We considered two variants of the EKT method: one using the canonical HF orbital EKT(HF) and another utilizing the variationally optimized pCCD orbitals — EKT(pCCD). 
All the implementations are available in the developer version of the PyBEST software package~\cite{pybest-paper-cpc-2021, pybest-paper-cpc-2024} and will be an official part of the next major release. 

The developed EKT(HF/pCCD) models were systematically benchmarked against eight neutral atoms and two molecular data sets: six small molecules depicted in Figure~\ref{fig:molecular-structures-2} and 24 organic acceptor molecules shown in Figure~\ref{fig:molecular-structures-1}.
In our tests, we employed cc-pVDZ, cc-pVTZ, aug-cc-pVDZ, and aug-cc-pVTZ basis sets. 
We found that EKT(HF) suffers from numerical instabilities when combined with small basis sizes, such as cc-pVDZ and cc-pVTZ, across the investigated test sets. 
The EKT(pCCD) is free of such issues and provides reliable IPs across all data sets and basis sets. 
The accuracy of EKT(pCCD) approaches that of computationally more expensive models, such as IP-EOM-fpCCD or CCSD(T).
The unparalleled advantage of the EKT(pCCD) approach is its ability to provide reliable IPs compared to experimental data, even with a small basis set size, such as cc-pVDZ.
Opposed to standard electronic structure methods, the quality of computed IPs within the EKT(pCCD) framework is insensitive to the presence of augmented functions in the basis set.
This low sensitivity to the inclusion of diffuse (augmented) functions in the basis set is a key practical advantage, which can be ascribed to the orbital optimization within pCCD.
That, combined with the low computational cost of pCCD ($\mathcal{O}(N^4)$) if the orbitals are optimized and/or the Cholesky representation of the electron repulsion integrals is used), offers a powerful tool for reliable IP predictions in large molecular assemblies. 
To that end, the EKT(pCCD) approach is extremely valuable in predicting the IPs of large organic molecules, including the high-throughput screening of novel organic compounds with desired properties.
On the other hand, an accurate description of EAs is more challenging. Our initial efforts of extending the EKT to predict EAs yield larger errors than those obtained within EA-EOM-pCCD.
An alternative route we will investigate in the future is the EKT formulation developed by Cioslowski and co-workers\cite{ekt-cioslowski-jcp-1997} to determine whether it provides improved EAs when applied within the pCCD framework.


\begin{acknowledgement}
S.~J., and P.~T.~acknowledge financial support from the SONATA BIS research grant from the National Science Centre, Poland (Grant No. 2021/42/E/ST4/00302). 
Funded/Co-funded by the European Union (ERC, DRESSED-pCCD, 101077420).
Views and opinions expressed are, however, those of the author(s) only and do not necessarily reflect those of the European Union or the European Research Council. Neither the European Union nor the granting authority can be held responsible for them. 
\end{acknowledgement}

\begin{suppinfo}
The following files are available free of charge:
\begin{itemize}
  \item si.pdf: Equations used for the statistical analysis, ionization potentials obtained with different thresholds for the extended Koopmans' method applied to atoms using various basis set sizes and molecular orbitals (both pCCD and HF). It also contains equations and IPs from Koopmans flavor models, as well as IP-EOM-CC energies for acceptor molecules with various basis sets and molecular orbitals (both pCCD and HF). Additionally, it includes xyz coordinates for six small molecules.

  \item si.xlsx: IPs from various Koopmans' flavor models and IP-EOM-CC energies calculated with different basis sets for eight atoms, six small molecules, and 24 acceptor molecules.

\end{itemize}
\end{suppinfo}

\section*{Data Availability Statements}
The data underlying this study are available in the published article and its Supporting Information.
The released version of the PyBEST code is available on Zenodo at \url{https://zenodo.org/records/10069179} and on PyPI at \url{https://pypi.org/project/pybest/}.

\section*{Conflicts of Interest}
There are no conflicts to declare.

\bibliography{p}

\end{document}